\documentclass[iop]{emulateapj}

\usepackage{apjfonts}
\usepackage{bm}
\usepackage{CJK}
\usepackage{color}
 
\newcommand{\tP}{{\it The Payne}}

\shorttitle{the payne: ab initio fitting of stellar spectra}
\shortauthors{Ting et al.}

\begin{document}

\begin{CJK*}{UTF8}{gbsn}
\title{The Payne: self-consistent ab initio fitting of stellar spectra}
\author{Yuan-Sen Ting (丁源森)\altaffilmark{1,2,3,4,5**}, Charlie Conroy\altaffilmark{5}, Hans-Walter Rix\altaffilmark{6}, Phillip Cargile\altaffilmark{5}}
\altaffiltext{1}{Institute for Advanced Study, Princeton, NJ 08540, USA}
\altaffiltext{2}{Department of Astrophysical Sciences, Princeton University, Princeton, NJ 08544, USA}
\altaffiltext{3}{Observatories of the Carnegie Institution of Washington, 813 Santa Barbara Street, Pasadena, CA 91101, USA}
\altaffiltext{4}{Research School of Astronomy and Astrophysics, Australian National University, Cotter Road, ACT 2611, Canberra, Australia}
\altaffiltext{5}{Harvard--Smithsonian Center for Astrophysics, 60 Garden Street, Cambridge, MA 02138, USA}
\altaffiltext{6}{Max Planck Institute for Astronomy, K\"onigstuhl 17, D-69117 Heidelberg, Germany}
\altaffiltext{**}{Hubble Fellow}
\slugcomment{Submitted to ApJ}

%
%
%
%
%
%

\begin{abstract}
We present \tP, a general method for the precise and simultaneous determination of numerous stellar labels from observed spectra, based on fitting physical spectral models. \tP\, combines a number of important methodological aspects: it exploits the information from much of the available spectral range; it fits all labels (stellar parameters and element abundances) simultaneously; it uses spectral models, where the atmosphere structure and the radiative transport are consistently calculated to reflect the stellar labels. At its core \tP\, has an approach to accurate and precise interpolation and prediction of the spectrum in high-dimensional label-space, which is flexible and robust, yet based on only a moderate number of {\it ab initio} models ($\mathcal{O}(1000)$ for 25 labels). With a simple neural-net-like functional form and a suitable choice of training labels, this interpolation yields a spectral flux prediction good to $10^{-3}$ {\it rms} across a wide range of $T_{\rm eff}$ and $\log g$ (including dwarfs and giants). We illustrate the power of this approach by applying it to the APOGEE DR14 data set, drawing on Kurucz models with recently improved line lists (Cargile et al., {\it in prep.}): without recalibration, we obtain physically sensible stellar parameters as well as 15 element abundances that appear to be more precise than the published APOGEE DR14 values. In short, The Payne is an approach that for the first time combines all these key ingredients, necessary for progress towards optimal modelling of survey spectra; and it leads to both precise and accurate estimates of stellar labels, based on physical models and without `re-calibration'. Both the codes\footnote{\href{https://github.com/tingyuansen/The_Payne}{https://github.com/tingyuansen/The\_Payne}} and catalog are made publicly available online.
\end{abstract}


%
%
%
%
%
%

\section{Introduction}
\label{sec:introduction}

Large-scale multiplexing spectroscopic surveys are revolutionizing the quality and quantity of spectroscopic data for Galactic archaeology. Surveys such as APOGEE \citep{maj17}, GALAH \citep{des15} and Gaia-ESO \citep{smi14} are collecting high-quality spectra for $10^5-10^6$ stars with a spectral resolution $R \simeq 25$,$000$, orders of magnitudes more stars than previous samples. Lower-resolution spectroscopic surveys, e.g., RAVE \citep{ste06}, Gaia-RVS \citep{rec16}, and LAMOST \citep{luo15}, are collecting even larger samples. And upcoming spectroscopic surveys, such as DESI \citep{desi16}, 4MOST \citep{dej14}, WEAVE \citep{dal16}, MOONS \citep{cir14}, SDSS-V \citep{kol17}, will boost sample sizes at both high and low spectral resolution by another order of magnitude, towards $\sim 10^7$ stars. 

However, learning about Galactic archaeology and stellar physics from these spectra depends crucially on our ability to correctly and precisely infer numerous stellar labels from these spectra: stellar parameters and individual elemental abundances. This requires a rigorous method to extract the maximal information from these data, based on physical {\it ab initio} spectral models. This is the focus of this study. 

A key to rigorous fitting of stellar spectra is the ability to fit all stellar labels (typically $>20-50$ for stellar spectra) simultaneously \citep{tin16b,rix16}, principally for two reasons: the spectral features of many elements are blended in the spectrum, imprinting a covariant signature on the data. And for quite a number of elements, variations in their abundances not only affect the strength of their spectral features, but also alter the stellar atmospheric structure \citep{tin16b}; this in turn affects the spectral features of other elements, especially in cooler stars. Therefore, spectral modeling should be based on {\it self-consistently} calculated models that take into account the dependence of the atmosphere structure on various element abundances. This dependence is widely implemented for changes in [Fe/H], but not other elements.

\begin{figure*}
\centering
\includegraphics[width=1.0\textwidth]{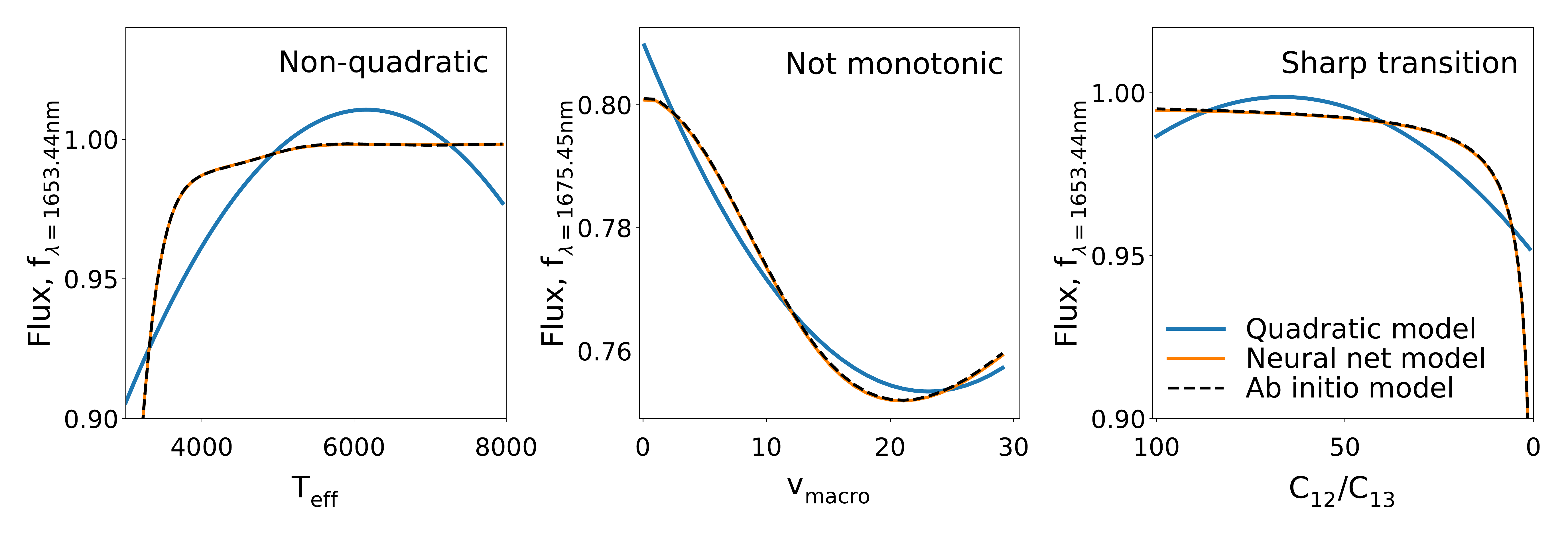}
\caption{High-fidelity spectral flux interpolation and prediction is enabled by \tP, compared to a quadratic flux model. In each of the three panels, the dashed line shows the expected flux variation of individual pixels with different label variations. \tP\, approximates these flux variation through neural networks (orange line), which are more flexible than quadratic models (blue line) in approximating the flux variation across a wide label range, without incurring much additional cost in {\it ab initio} model calculation for fitting a spectrum. The three panels show three different scenarios where quadratic models do not approximate the flux well. (a) A spectral region with strong sensitivity to $T_{\rm eff}$. (b) A multiply non-monotonic variation of flux, e.g., with $v_{\rm macro}$. (c) A transition where a label changes abruptly and has little effect below a threshold, e.g., with $C_{12}/C_{13}$. In contrast, \tP\, (solid orange line) has no problem approximating these variations.}
\label{fig1}
\end{figure*}

In practice, current spectral analyses often fit only small portions of the spectrum to determine any particular element abundance, holding the abundances of other elements fixed. And they often require subsequent recalibration of the basic stellar parameters (e.g., $\log g$ and $T_{\rm eff}$) or abundance-$T_{\rm eff}$ trends inferred from the spectral fitting. This motivates the need for the development of a comprehensive approach to study these issues. Here we will present such a method, \tP\,\footnote{In appreciation of Cecilia Payne-Gaposchkin's ground-breaking work on physical spectral models.} in this study.

\tP\, combines a number of important ingredients: a set of spectral models based on a state-of-the-art line list (Cargile et al., {\it in prep.}); models computed that are self-consistently calculated for each set of labels; a robust and flexible ``interpolator'' in the high-dimensional label space for spectral fitting that can precisely predict spectral model fluxes for arbitrary sets of labels; a well-defined and objective assessment and mitigation of the wavelength regions where the models have important systematic shortcomings; and a robust estimate of the label estimates from the entire remaining parts of the observed spectra. For modeling stellar spectra, \tP\, is a fully automated, simple, transparent fitting machinery, given a set of {\it ab initio} synthetic spectral models. The codes for running \tP\, are publicly available online. Moreover, the fitting is very efficient -- e.g., fitting 25 labels for an APOGEE spectrum with \tP\, takes less than 1 CPU second. \tP\, differs from {\it The Cannon} \citep{nes15a,cas16} principally in two respects: it is based on physical instead of data-driven models, and it generalizes the ``interpolator'' beyond the quadratic polynomial implemented in {\it The Cannon} and \citet{rix16}. In short, The Payne is an approach that for the first time combines all these ingredients, necessary for progress towards optimal modelling of survey spectra; and it leads to both precise and accurate estimates of stellar labels, based on physical models and without `re-calibration'.

This paper is structured as follows: we introduce \tP\, and test the interpolator at its core in Section~\ref{sec:introducing-the-payne}. We apply \tP\, to the APOGEE DR14 data set in Section~\ref{sec:implementing-apogee}, and present the resulting catalog. We discuss the outlook of stellar spectroscopy in the light of \tP\, in Section \ref{sec:discussion} and conclude in Section~\ref{sec:conclusion}.

%
%
%
%
%
%

\section{The Payne}
\label{sec:introducing-the-payne}

%
%
%
%
%
%

\subsection{Motivation}

Current approaches to modeling stellar spectra, either with physical or data-driven models, have important limitations that are well-documented in the recent literature \citep{boe11,adi12,ben14,blan14,nis14,hol15,nes15a,boe16,cas16,gar16,rix16,tin16b,tin17a,tin17b,zha16,elb18a,elb18b}. In this section we present our approach to addressing some of these limitations\footnote{For example, to fully harness the information from spectra, a full spectral fitting method can be advantageous \citep[see detailed discussion in][]{tin16b,tin17a} than equivalent width based methods, as much of the spectral information is embedded in the subtle blended features.}. At the core of \tP\, is the ability to perform full simultaneous spectral fitting of all stellar labels through an efficient but precise way of ``interpolating'' a modest set of synthetic model spectra in high-dimensional label space. 

The key idea for efficiently interpolating an ensemble of synthetic models is two-fold. First, we do not need to create a high-dimensional ``grid'' of model spectra, which would be computationally prohibitive for, say, 25 labels in this study; with an adaptive approach described below we only create models within the label space spanned by the data and ``where needed''. Second, we resort to building a generative model for the spectra at arbitrary point (in a portion) of label space, as \citet{nes15a} and \citet{rix16}. If the model for the spectral flux at each pixel is forced to be a quadratic function of the $N$ labels, then only a few times $N \times (N+1)/2$ {\it ab initio} spectral models are needed as a basis.

While quadratic models are simple and elegant, they limit the portion of label space over which precise ($\sim 10^{-3}$) flux predictions are possible. For fitting a broad range of stellar labels (e.g., fitting dwarfs and giants or $3000~{\rm K} \le T_{\rm eff} \le 8000~{\rm K}$ simultaneously), quadratic flux models appear too restrictive. Furthermore, for stellar labels such as $v_{\rm macro}$, at any given pixel, the variation of flux can be more complicated and often not monotonic. Such complex label-dependences of the flux are illustrated by three one dimensional examples in Fig. \ref{fig1}. In this figure we show the dependence of continuum-normalized flux as a function of $T_{\rm eff}$, $v_{\rm macro}$, and $C_{12}/C_{13}$. Here we assume the same Kurucz synthetic models as we will describe in Section~\ref{sec:synthetic-models} convolved with the APOGEE averaged line spread function (LSF) to simulate the variations we expect from APOGEE. Clearly a quadratic model cannot capture the behavior of the flux over the entire parameter range, while a more flexible neural network can reproduce the variation in the model very well, as we quantify in greater detail below.

\begin{figure*}
\centering
\includegraphics[width=1.0\textwidth]{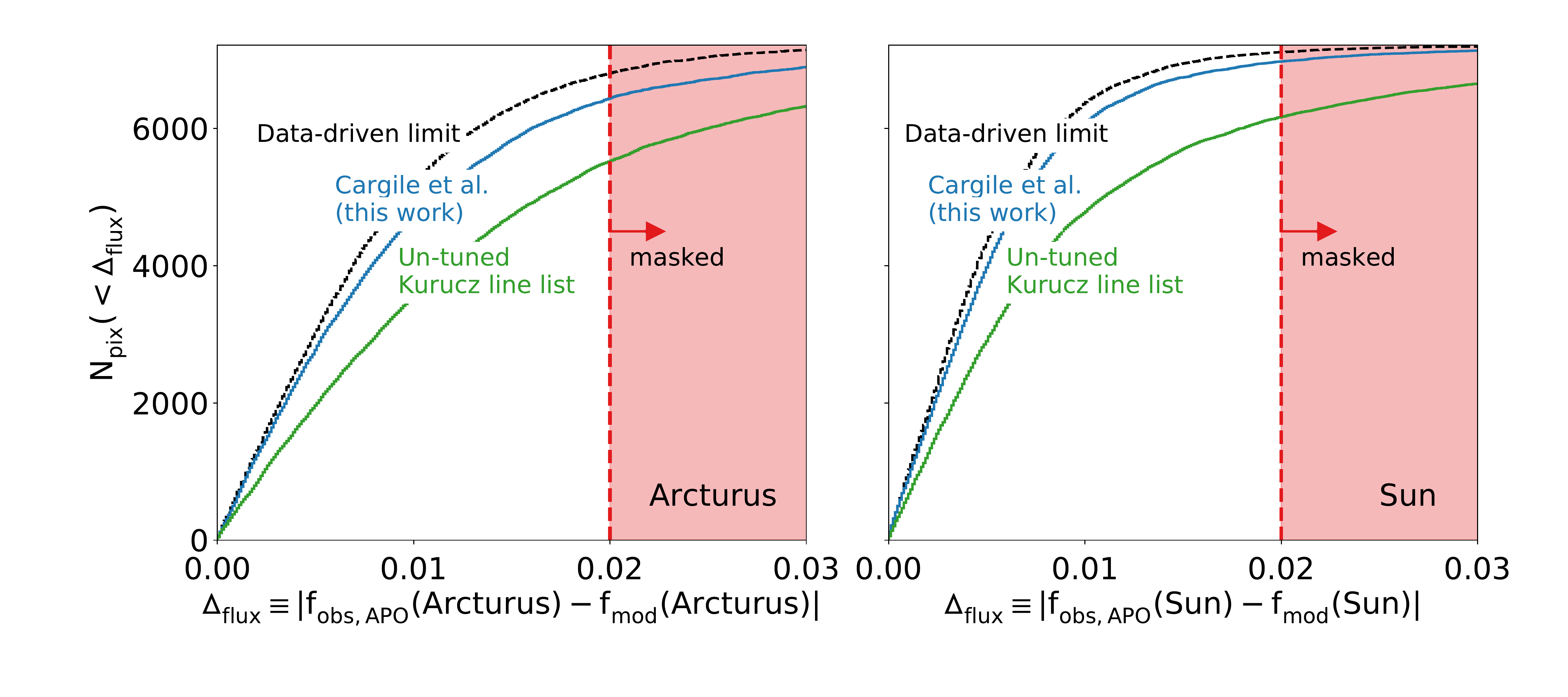}
\caption{Comparison of the model quality for the improved line list (Cargile et al., {\it in prep.}) adopted in this study versus an un-tuned Kurucz line list. We generated synthetic models (with those two line lists) adopting fiducial stellar labels for Arcturus and the Sun. These models were convolved with the APOGEE-determined average LSF and compared to the APOGEE spectra of Arcturus and the Sun. The panels show the cumulative distribution of APOGEE wavelength pixels as a function of the absolute deviation of the models and the observations. For comparison, we also convolved the FTS spectra of Arcturus and the Sun, observed at very high-resolution ($R=300$,$000$) and high S/N, with the same LSF and compared those to the corresponding APOGEE spectra (dashed black lines). This comparison reflects the limit of a perfect model. Due to the influence of telluric lines, an imperfect LSF, and perhaps other data-related systematics, the convolved FTS spectra do not exactly match the observed APOGEE spectra. The vertical dashed line shows the model-data mismatch threshold that we adopt in this study for creating a pixel mask for the fitting procedure. Pixels more discrepant than this cut in either the Sun or Arcturus are omitted. The improved line list allows us to discard far less spectral information, and hence improve the precision of our fit.}
\label{fig2}
\end{figure*}

%
%
%
%
%
%

\subsection{Neural networks for precise model spectrum prediction}
\label{sec:applying-neural-network}

The interpolation and approximation of functions with neural networks is based on the idea that most functions can be approximated by a finite composite of simple functions. For \tP\, we consider an extremely simple neural network architecture, ``fully-connected'' with only two hidden layers. At each wavelength pixel $\lambda$, we posit that the flux as a function of stellar label $\boldsymbol{\ell}$, can be written as
\begin{equation}
f_\lambda = w \cdot \sigma\bigg( \tilde{w}_\lambda^i \sigma \Big( w^k_{\lambda i} \ell_k + b_{\lambda i} \Big) + \tilde{b} \bigg) + \bar{f}_\lambda,
\label{eq:the-payne}
\end{equation}

\noindent
where $\sigma$ is the Sigmoid function $\sigma (x) = 1/(1 + e^{-x})$, and where we have assumed the Einstein convention for index summation. In the training step, we seek the coefficients $(w, \tilde{w}_\lambda^i, \tilde{b}, w^k_{\lambda i}, b_{\lambda i}, \bar{f}_\lambda)$ that best approximate the training spectra as a function of their stellar labels in the least squares sense. This formalism can be viewed as a straightforward extension of the quadratic flux models \citep[e.g.,][]{nes15a,rix16}, which essentially adopts the flux model $w^i \ell_i + w^{jk} \ell_j \ell_k$. In \tP\, we consider the non-linear composite function $w \cdot \sigma ( \tilde{w}_\lambda^i \sigma ( w^k_{\lambda i} \ell_k + b_{\lambda i}) + \tilde{b})$ to be the more flexible expansion terms. The rationale for adopting a more flexible functional form is similar to taking a higher ``expansion'' order such that the ``Taylor'' expansion convergence sphere encompasses a larger region of the parameter space of interest \citep[see][for a more detailed discussion]{tin16b}.

The number of ``neurons" $i$ in Eq.~\ref{eq:the-payne} is a free hyper-parameter to be optimized. Increasing the number of neurons enables the approximation a more complicated function, but at the risk of over-fitting the function. Besides adopting more number of neurons, one can also increase the complexity of the neural networks by increasing the number of ``layers'' by considering the composite of the current composite functions -- i.e., $f_\lambda \sim \sigma ( \cdots \sigma ( \cdots \sigma (\cdots ))$

Cross-validation experiments described below motivate the following choices. We adopt a two hidden layers model with 10 neurons. This choice was initially motivated by the fact that the number of free coefficients in this simple neural network model is comparable to that in a quadratic model. At least for stellar spectra, designing the neural networks to have roughly the same number of coefficients of simple polynomials seems to be a robust practical guideline. We checked that adopting a significantly more complex neural network model does not improve the qualitative results of this study, but does lead to over-fitting. We train the neural networks by minimizing the $L^2$ loss, i.e., minimizing the sum of the Euclidean distances between the target ({\it ab initio} flux and the model-predicted (or, ``interpolated'') flux at each pixel. We found no need for further imposing explicit $L^1$ regularization \citep[e.g.,][]{cas16} to the networks as it does not improve the results presented in this study. We limit ourselves to small networks precisely to avoid overfitting, as such regularization is not necessary.
\begin{figure*}
\centering
\includegraphics[width=1.0\textwidth]{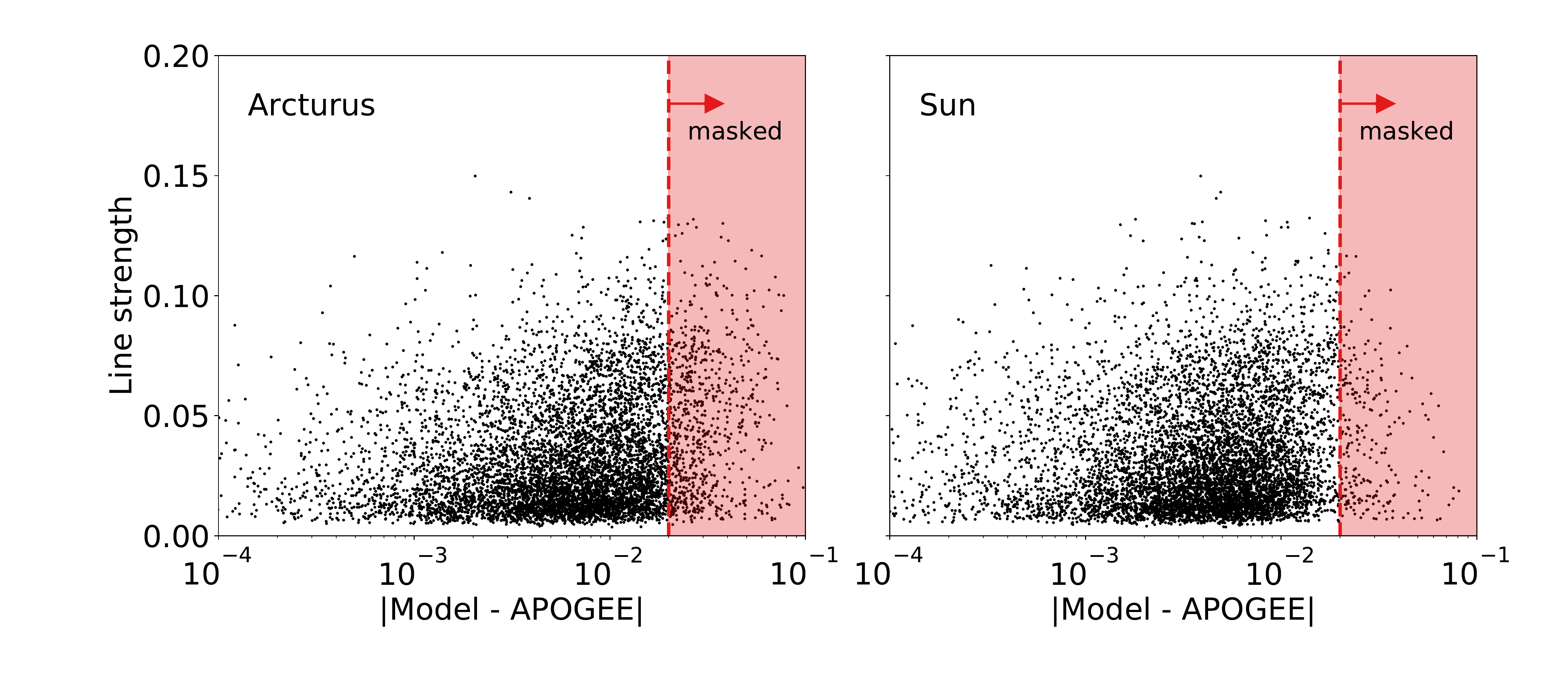}
\caption{Most of the strongly varying spectral features are retained in \tP's fitting mask. Analogous to Fig.~\ref{fig2}, the $x$-axis shows the deviation of the model from the Arcturus and Solar spectra observed by APOGEE at the APOGEE resolution. The $y$-axis shows the normalized flux {\it rms} for individual pixels, among synthetic training spectra of \tP. Larger $y$-axis values indicate that the pixel contains a strongly varying spectral feature; those are the pixels that can discriminate between models. There is an overall weak correlation between the model deviation with the feature strength because stronger broad features could be harder to model. Nonetheless, with the pixel mask we apply in this study, most strong features remain included in the fit, and we only discard a modest amount of spectral information.}
\label{fig3}
\end{figure*}

Neural networks are of course not the only flexible model ``interpolators'', as Gaussian processes or support vector regressions are also employed in related circumstances. For the case at hand, \tP\, has the advantage of being much faster computationally. While the training of neural networks is more computationally expensive than the quadratic models (each wavelength pixel takes about 5 CPU minutes), once the neural networks are trained, the speed of inference is about same as the quadratic models, and is independent of the training set size, as we simply need to evaluate the composite functions. While Gaussian processes are powerful for full Bayesian inferences, predicting a model spectrum at a new label point through Gaussian processes can be extremely slow: it requires the inversion of a matrix, has a complexity of $\mathcal{O}(N_{\rm train}^3)$, and can be very memory intensive. 

Finally, the fundamental idea of \tP\, is different from some of other previous applications of neural networks in spectral analyses \citep{fab18,leu19}. These studies attempted to map spectrum to the stellar label through neural networks, but in this study, we are mapping stellar label to the spectrum. Summarizing the detailed pros and cons of these methods are beyond the scope of this study; here we will only briefly discuss the logic behind our choice. Direct mapping from spectrum to stellar label can be advantageous as the spectral-fitting component becomes trivial -- evaluating stellar labels in this case only requires evaluating the mapping/function directly, which is extremely fast. On the other hand, mapping $f:$ spectrum $\rightarrow$ label limits the ability to differentiate the function with respect to the label, unlike \tP\,, which has $f:$ label $\rightarrow$ spectrum. Differentiating the emulating function with respect to label can be useful in many cases -- especially at low-resolution, comparing $\partial f/\partial$label to theoretical line lists can be the key to enforcing that elemental abundance are derived from their corresponding absorption features instead of astrophysical correlations. It also allows us to impose theoretical prior as was done in \citet{tin17b} \citep[but see][]{leu19}. This reason prompted our choice to map from stellar label to spectrum \citep[see also][]{daf16}. The downside of this approach, however, is that evaluating the label requires least squares minimizations, which is slower than simply evaluating a function. In short, both types of mapping have their own merits, and which method to use clearly depends on the applications.
 
\begin{table*}
\begin{center}
\caption{Sampling scheme for the model grid in this study.\label{table1}}
\begin{tabular}{lcc}
\tableline \tableline
\\[-0.2cm]
Stellar label & Sparse grid & Refined grid \\[0.1cm]
\tableline
\\[-0.2cm]
$n_{\rm grid}$ & 2000 & The same \\[0.1cm]
$T_{\rm eff}$, $\log g$ & \begin{tabular}{c} Draw uniformly from the MIST isochrone convex hull \\ with $T_{\rm eff} \in [3000,8000]\,$K, $\tau_{\rm age} \in [3,10]\,$Gyr \end{tabular}  & The same \\[0.2cm]
${\rm [}$Fe/H] & Draw uniformly, ${\rm [Fe/H]} \in [-1.5,0.5]$  & The same \\[0.1cm]
$v_{\rm micro}$ & Draw uniformly, $ v_{\rm micro} \in [0.1,3] \, {\rm km/s}$  & The same   \\[0.1cm]
$v_{\rm macro}$ & Draw uniformly, $ v_{\rm macro} \in [0,30] \, {\rm km/s}$  & The same   \\[0.1cm]
$C_{\rm 12}/C_{\rm 13}$ & $f_{\rm MIST}(T_{\rm eff}, \log g, {\rm [Fe/H]})$ + a scatter of $\pm 35$ & The same  \\[0.1cm]
$[X/{\rm Fe}]$ & Draw uniformly, $[X/{\rm Fe}] \in [-0.5,0.5]$ & $P([X/{\rm H]|[Fe/H]})$ from the sparse grid APOGEE fits, with $\Delta {\rm [Fe/H]} = 0.2\,$dex \\[0.1cm]
\tableline\\[0.2cm]
\end{tabular}
\end{center}
\end{table*}

%
%
%
%
%
%

\subsection{The choice of stellar training labels for spectral model building}
\label{sec:priors}

Beyond the choice of how to interpolate among a set of model grid points, another essential choice must be made: the training set size and the stellar labels at which the {\it ab initio} models are to be evaluated to provide training spectra. Formally, we require barely more training spectra than the number of free parameters in the neural networks, which would be 273 training spectra in the case at hand. However, uniformly distributing few hundreds training labels in a high dimensional ($N_{\rm dim} = 25$) space would not be optimal because the distribution will be too sparse in the label space, and the interpolation will not be precise. But in generative models like \tP\, we need not draw from a regular, uniformly spaced training labels.

As discussed in \citet{tin15}, generating training spectra around the label space where real observed stars are expected to occupy can exponentially reduce the number of models needed. The volume of a hyper-ellipsoid in a high dimensional space is exponentially smaller than the volume of a hypercube where the training labels are uniformly distributed. In our illustrative application of \tP, we fit 25 stellar labels, including all elemental abundance with entries in our line list within the APOGEE spectral range. As stellar parameters, we fit $T_{\rm eff}$, $\log g$, $v_{\rm micro}$, $v_{\rm macro}$, and $C_{\rm 12}/C_{\rm 13}$ along with the 20 elemental abundances (C, N, O, Na, Mg, Al, Si, P, S, K, Ca, Ti, V, Cr, Mn, Fe, Co, Ni, Cu, Ge). We consider a training set of 2000 training spectra. \citet{rix16} showed that adopting more training set than the free parameters will better constrain the flux variation, especially when the range of the parameter space is large. We found that adopting a 10 times larger training set does not change our results qualitatively in this study. For the 2000 training spectra, we adopt an adaptive refinement technique to decide on the training labels as described below.

We start with a ``sparse'' set of labels that samples $T_{\rm eff}$ and $\log g$ from the MIST isochrones \citep{cho16} assuming $[Z/{\rm H}] = -1.5$ to 0.5, $T_{\rm eff} = 3$,$000-8$,$000\,$K and stellar age from $3-10$ Gyrs, covering both dwarfs and giants. We consider stellar evolution states from the main sequence to the core helium burning at the red clump. We then use these labels to create two convex hulls for the giants (defined with $\log g < 4$) and the dwarfs ($\log g > 4$) separately in the $T_{\rm eff} - \log g$ space, i.e., minimum polygons that encompass the tracks from the MIST isochrones. Subsequently, we randomly sample $T_{\rm eff}$ and $\log g$ from a uniform distribution within these convex hulls. Analogously, we draw $v_{\rm micro}$ uniformly from $0.1-3\,$km/s and $v_{\rm macro}$ uniformly from $0-30\,$km/s with 2000 data points. We have found that this choice spans most of the derived APOGEE label space without requiring extrapolation. For $C_{12}/C_{13}$ we assume a weak prior. We adopt the isochrones value of $C_{12}/C_{13}$ given the stellar parameters of the training data. But we arbitrary spread out the $C_{12}/C_{13}$ values on the training set with a uniform distribution of $\pm 35$. Finally, for this sparse grid, we randomly draw all elemental abundances $[X/{\rm H}]$ from a uniform distribution with the condition $-0.5 < [X/{\rm Fe}] < 0.5$. Note that, here we train a single spectral model that encompasses both dwarfs and giants.

While the sparse grid is essential to make sure that we capture all cases, spanning a 25-dimensional space with only 2000 training data cannot constrain the flux variation to the needed precision. Therefore, we need to refine the label space from which we draw our training labels. To do that we train \tP\, with the sparse grid, fit all APOGEE spectra, which results in an initial distribution of the sample in label-space. Then, we re-sample 2000 training data points with $[X/{\rm H}]$ drawn from these initial APOGEE label values. We note that APOGEE data does not span the $T_{\rm eff}-\log g-{\rm [Fe/H]}$ space uniformly. Therefore, to avoid only fitting the variation of flux well at the at the bulk of the data, we do not resample main stellar parameters with the fitted values, but rather we sample $T_{\rm eff}$, $\log g$, [Fe/H], $v_{\rm micro}$, $v_{\rm macro}$ and $C_{12}/C_{13}$ as before. But we adopt $[X/{\rm H}]$ from the fitted APOGEE values that have consistent ${\rm [Fe/H]}$. In other words, we bin the measured (using \tP\, trained on the sparse grid) $[X/{\rm H]}$ APOGEE values according to their fitted [Fe/H] values with a bin size of 0.2 dex. We only sample $[X/{\rm H}]$ in the corresponding [Fe/H] bin consistent with the newly drawn [Fe/H] training label. And these 2000 resampled training points constitute the final training set. Our sampling scheme is summarized in Table~\ref{table1}.
\begin{figure*}
\centering
\includegraphics[width=1.0\textwidth]{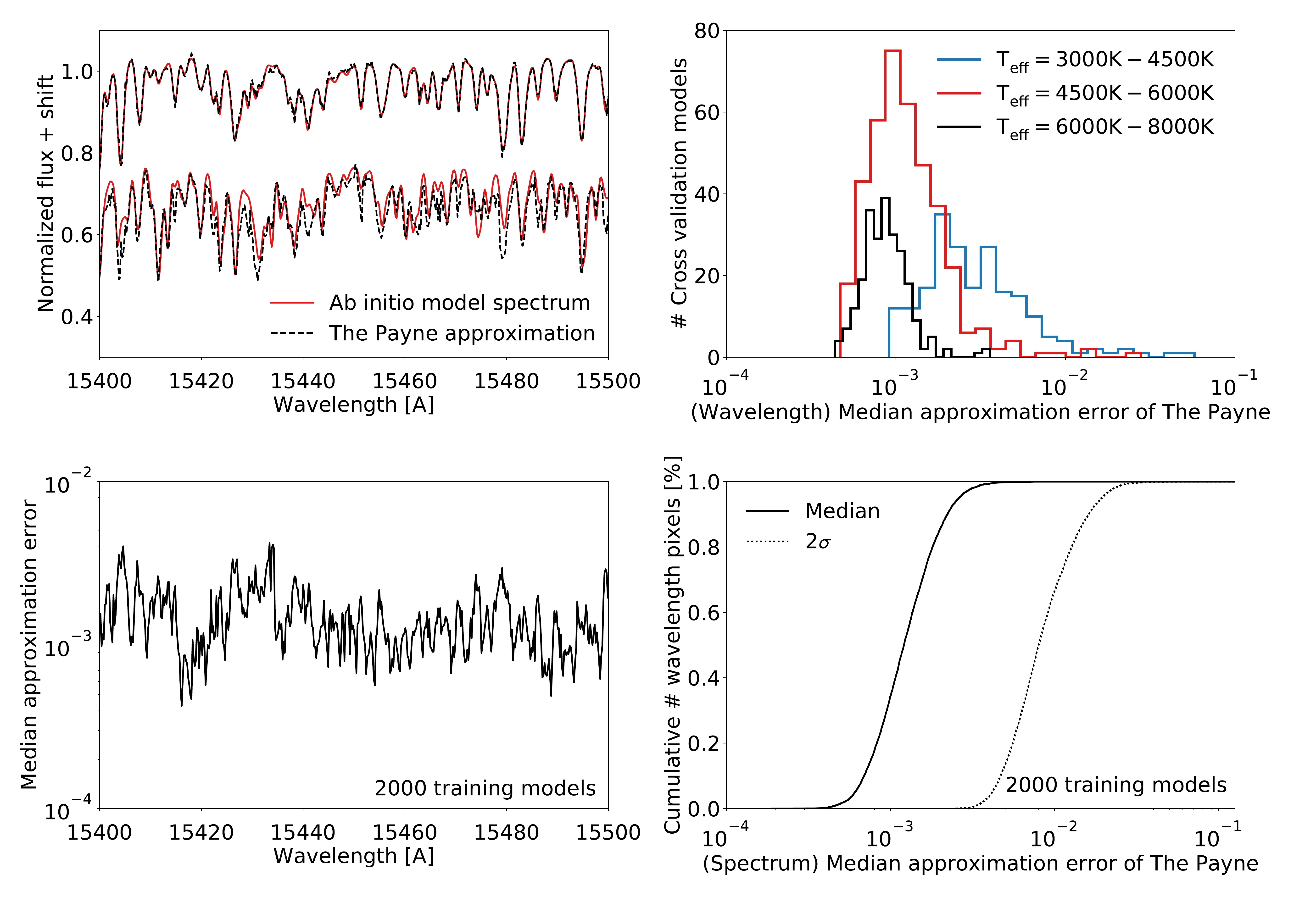}
\caption{Assessing the flux prediction quality of \tP. For 25 labels, we have adopted 2000 {\it ab initio} models as the training set to build the spectral model, and have used another 850 {\it ab initio} models for cross-validation. The top left panel shows two examples of spectral predictions from \tP\, compared to the directly calculated cross-validation spectra: the upper spectrum show a case where \tP\, prediction and cross-validation spectrum appear indistinguishable. Most of the validation spectra are in this category. The lower spectrum show an extreme case where the interpolation is poor (errors $> 1\%$). We shift the continuum baseline of the lower spectrum for the clarity of the plot. The top right panel shows a more quantitative assessment of how the quality of \tP's approximate flux prediction varies across label space. We calculate the median of approximation errors of individual testing spectra over all wavelength pixels. Cooler stars have more strong transitions, and hence they are harder to model and interpolate; but even for the cooler case, the median approximate is less than 1\%, which is smaller than the typical S/N of an observed APOGEE spectrum. The lower panels demonstrate the median approximate error of individual wavelength pixels over all testing spectra. The lower left panel illustrates the median approximation error for (across the ensemble of cross-validation spectra) a small segment of the wavelength range, and the lower right panel shows the cumulative distribution for all wavelength pixels. \tP\, approximates the flux (variation) for each pixel to the level of $\sim 0.1\%$.}
\label{fig4}
\end{figure*}

\begin{figure*}
\centering
\includegraphics[width=1.0\textwidth]{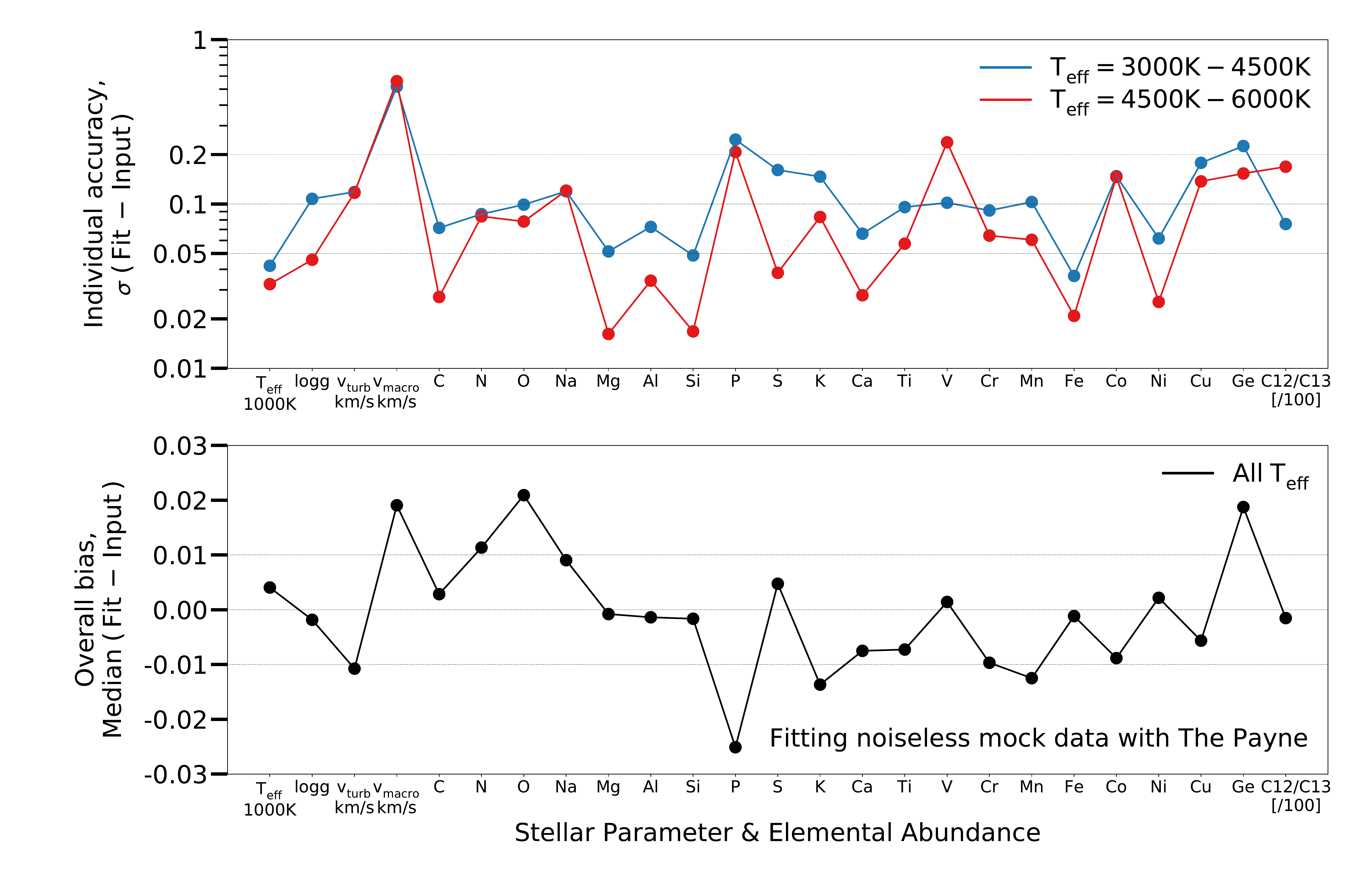}
\caption{Recovery of stellar labels when applying \tP\, to fitting noiseless {\it ab initio} cross-validation spectra. Shown in the top panel is the deviation of the recovered labels from the input labels which reflect the systematic flux errors, incurred in interpolation with \tP. This sets the floor for the stellar label {\em accuracy} (not precision, see text for details). The red line shows deviation for the hotter stars and the blue line for the cooler stars. For individual stars, we might incur a bias of $0.03\,$ dex in [Fe/H], $0.03-0.1\,$dex in other elemental abundances (C, N, O, Mg, Al, Si, S, K, Ca, Ti, Cr, Mn, Fe, Ni), $50\,$K in $T_{\rm eff}$ and $0.05\,$dex in $\log g$. Hotter stars have about two times smaller errors than these values because they are less subjected to interpolation error with \tP. In the bottom panel, we show the median deviation of the fit from the input labels. The bottom panel shows that, provided that the validation labels in the mock data are a fair representation of the APOGEE data, there is no strong global biases in the recovery.}
\label{fig5}
\end{figure*}

\begin{figure*}
\centering
\includegraphics[width=1.0\textwidth]{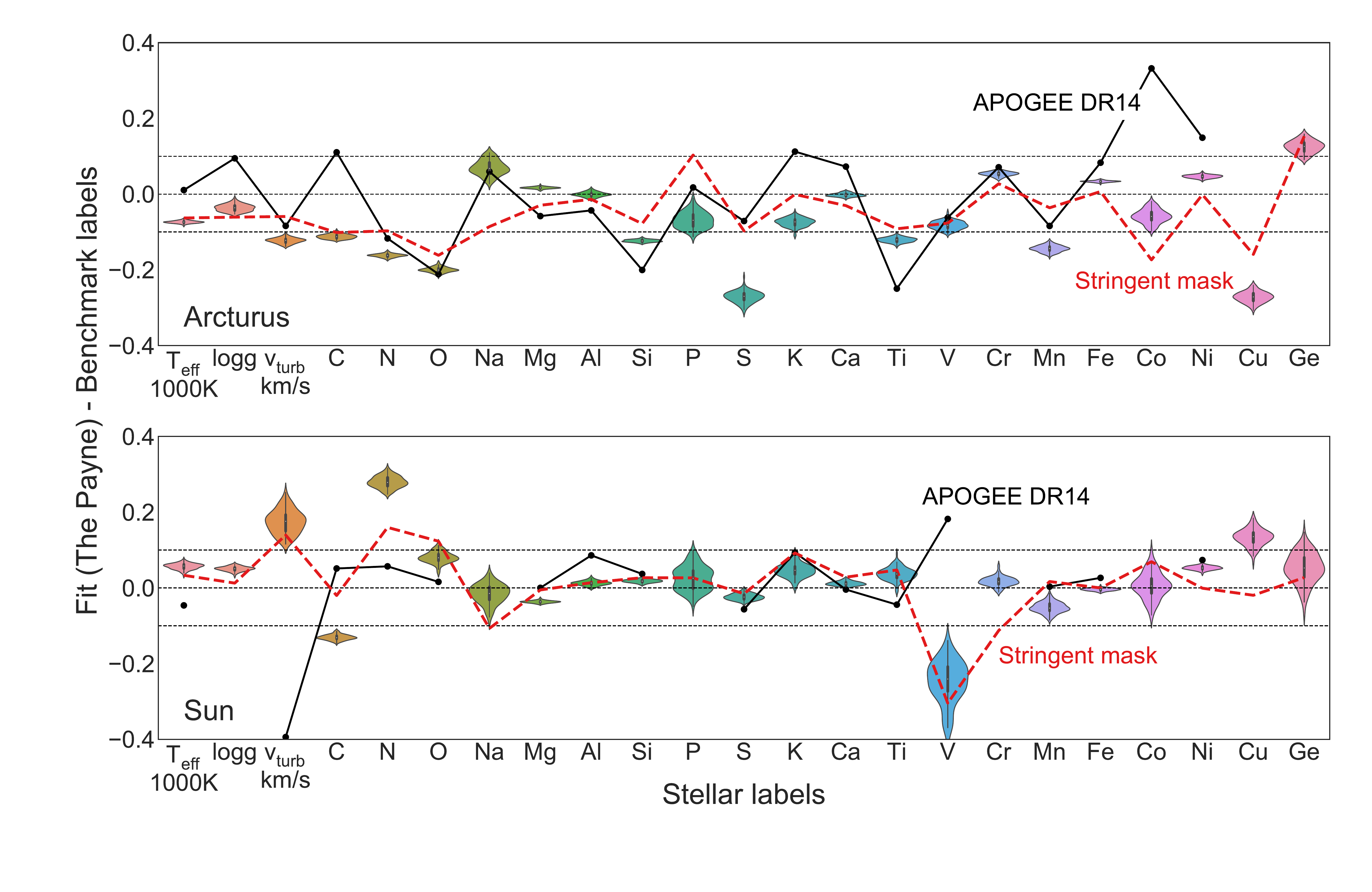}
\caption{Fitting APOGEE spectra of the Sun and Arcturus with \tP. We adopt the APOGEE Arcturus and Solar spectra and generate 100 realizations by sampling the reported uncertainties (S/N$\sim 400$). We fit these 100 realizations with \tP\, and, for the fiducial case with a spectroscopic mask of 2\% error tolerance, we plot the departure of the fitted values from the Arcturus and Solar reference stellar labels as the violin plots. The red dashed lines show the fit from \tP\, where a more stringent spectroscopic mask (0.5\% tolerance) is adopted. The APOGEE DR14 fitted values, when available, are overplotted with solid black lines as references. The fitted values are consistent with the reference values to about 0.1 dex in elemental abundances, similar to APOGEE DR14, with the exception of V at solar temperature because V only has an extremely weak feature at the Solar temperature. A more stringent spectroscopic mask reduces some biases, but at the expense of the precision for the overall sample.}
\label{fig6}
\end{figure*}

%
%
%
%
%
%

\subsection{The details: Ab initio models, line lists, spectroscopic masking, instrumental dispersion and continuum normalization}
\label{sec:synthetic-models}

We compute 1D LTE spectral models adopting the state-of-the-arts codes {\sc atlas12} and {\sc synthe} maintained by R. Kurucz \citep[][and reference therein]{kur70, kur81, kur93, kur05, kur13, kur17}. It is crucial to re-calculate the stellar atmospheric structure as we vary the stellar labels to obtain accurate stellar labels from APOGEE, instead of simply running the radiative transfer code. We calculate the stellar atmospheric structure by partitioning the stellar atmosphere into 80 zones of Rosseland optical depth, $\tau_R$, with the maximum Rosseland depth $\tau_R=1000$. When generating synthetic models, we automate the inspection of numerical convergence for each layer of the stellar atmospheres. We adopt Solar abundances from \citet{asp09} and the Arcturus stellar labels from \citet{ram11} throughout this study. We assume a standard mixing length theory with no overshooting for convection. After the stellar atmosphere converges, we produce the synthetic model spectra through the radiative transfer code {\sc synthe} at the nominal spectral resolution of $R=300$,$000$. The synthetic spectra are subsequently convolved to the APOGEE resolution assuming the APOGEE averaged LSF template. We normalize both the synthetic spectra and the APOGEE observed spectra following \citet{nes15a}. In this method, a set of wavelength pixels with the least response to stellar labels, based on the data-driven model {\it The Cannon}, are selected. A fourth order polynomial is fitted through the fluxes of these wavelength pixels and is used to approximate the continuum.

A crucial improvement of our {\it ab initio} models is the use of an updated line list (Cargile et al., {\it in prep.}), which will soon be made publicly available. Improving on the original Kurucz line list, the new line list tweaks three line parameters for every line stronger than 1\% at $R=300$,$000$ in either the Sun or Arcturus: the central wavelength, oscillator strength and the dominant broadening parameter. These line parameters are simultaneously fit to the high resolution spectral atlas of the Sun and Arcturus in several angstrom segments in order to capture possible covariance between overlapping lines. We refer readers to the paper for more details. Fig.~\ref{fig2} shows a comparative assessment of the new line list. We synthesize spectra at the Solar and Arcturus stellar labels, convolve and normalize them to the APOGEE resolution with the methods described above. We then compare the models to the observed Arcturus and Solar spectra from APOGEE. There is a total of 7214 pixels in an APOGEE spectrum, and Fig.~\ref{fig2} shows the cumulative number of wavelength pixels as a function of the absolute deviation of the models from the observations at each pixel.

The model-data match based on the updated line list adopted in this study is shown by the blue line, while the match with models that use the standard un-tuned line list (available on R. Kurucz's website) are shown by the green line. The shaded regions identify the pixels we mask and eliminate in the subsequent modeling -- pixels that have normalized model fluxes deviating by $\ge$2\% at the APOGEE resolution from the observed spectra, either for Arcturus or the Sun. About 90\% of pixels that we mask are due to disagreement with Arcturus especially in the middle chip of APOGEE, i.e., $15$,$800-16$,$400\,$\AA. The poorer agreement with Arcturus is not surprising because the line list is better calibrated to the Sun than to Arcturus, and because the cooler temperature of Arcturus results in more and stronger lines than in the Sun. The 2\% cut is chosen to produce a satisfactory balance between the accuracy and the precision of our derived stellar labels. Imposing a more stringent cut will minimize the systematic errors of the spectral models, but at the expense of the precision we can achieve because we are excluding more spectral information. Also note that this binary spectroscopic mask only discards $12\%$ of the APOGEE spectra, and we are still performing full spectral fitting with all stellar labels simultaneously. This should be distinguished from the ASPCAP mask which APOGEE DR14 imposed, where individual abundances are determined with different filters.

\begin{figure*}
\centering
\includegraphics[width=1.0\textwidth]{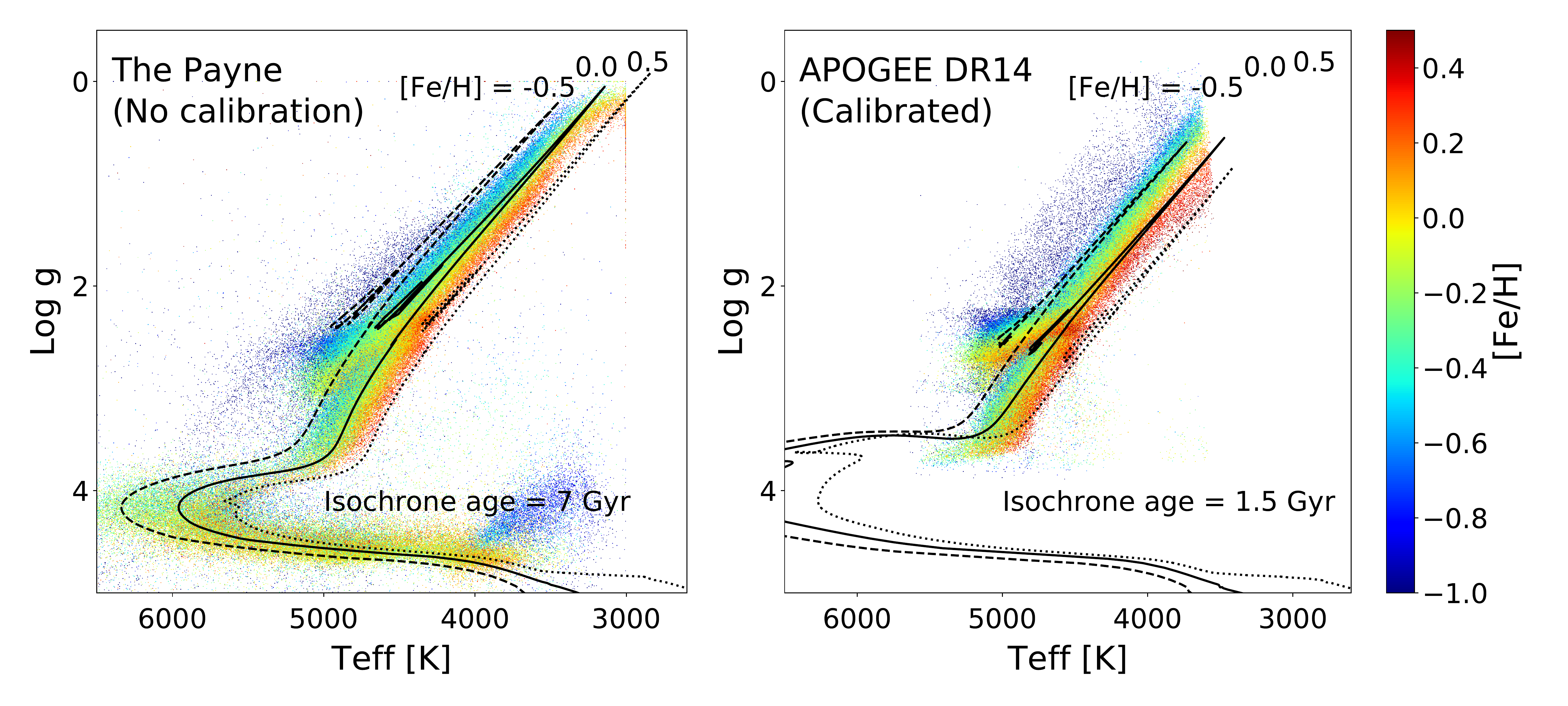}
\caption{ \tP\, measures physically sensible $T_{\rm eff}$, $\log g$ and [Fe/H] for both giants and dwarfs simultaneously without requiring external calibration. On the left-hand side, we show \tP\, $T_{\rm eff}$-$\log g$ Kiel diagram overplotted with MIST isochrones assuming a stellar age of 7 Gyr. On the right-hand side, we show the APOGEE DR14 {\em calibrated} counterparts and with MIST isochrones at 1.5 Gyr. \tP\, derives stellar parameters that are consistent with stellar isochrones for both giants and dwarfs star with only a single \tP\, model. For metal poor dwarfs with $T_{\rm eff} < 4000\,$K, the results deviate strongly from the isochrones. This could be due to the Kurucz models 1D stellar atmosphere is a poor assumption in this regime or simply the line list is not well calibrated at this temperature and metallicity range.}
\label{fig7}
\end{figure*}

Fig.~\ref{fig2} shows the comparisons of the APOGEE spectra of the Sun and Arcturus, compared to the convolved version of the very high S/N resolution $R=300,000$ FTS spectra of the Sun and Arcturus, serving as ``perfect model" templates. The convolved high-resolution observed Solar and Arcturus spectra do not match their APOGEE counterparts perfectly for several reasons. The APOGEE H-band suffers from severe telluric contamination which is imperfectly subtracted. Furthermore, the LSF and continuum normalization that we adopt are not perfect and could contribute to this discrepancy. Nonetheless, the convolved FTS spectra set the baseline for the best case scenario and show that the updated line list is closer to this limit than the original Kurucz line list. We also tested that making a spectroscopic mask at the FTS resolution and subsequently convolving it to the APOGEE resolution does not work. For \tP, it is crucial to make the spectroscopic mask directly in the observable space. The mask is meant to capture both for theoretical imperfections (imperfect line parameters, non-LTE effects, etc.) and for observational problems (LSF, telluric absorption, etc.).

In Fig.~\ref{fig3} we further investigate which pixels are masked from the fit. The $y$-axis quantifies how informative each pixel is, quantified by the {\it rms} of the model variations when sampling the training labels; the $x$-axis shows the absolute deviation of the model from the observed spectrum for both Arcturus and the Sun. The {\it rms} is calculated with the refined synthetic model grid used in the final training. The shaded regions show pixels that are excluded from analysis. Fig.~\ref{fig3} shows that, overall, there is a weak correlation between the deviation and the spectral feature strength. This trend is expected because stronger lines are generally harder to model. But as shown, most of the spectral features are included in our fit, and only a minimal number of spectral features are masked.
 
Finally, we note that our method is completely general and can be applied to other spectroscopic models. We also tried to apply The Payne to the un-tuned Kurucz line list. We showed that, similar to the results using the new line list as we will present in this paper, the fits even with the old line list exhibit better agreements with the isochrones as well as a flat Teff-abundance trend for open clusters. However, the overall accuracy and precision with the old line list are not as good as the improved new line list. The worse precision is expected because, with the old line list, we need to mask out significantly more pixels (Fig.~\ref{fig2}). The slightly worse accuracy (i.e., not as good an agreement with the isochrones) is a bit puzzling. It suggests that the H-band spectroscopic models are not consistent throughout all the APOGEE pixels. Checking how the results vary by restricting to different sub-ranges of wavelength could shed light on this issue, but this is beyond the scope of this paper. Moreover, a thorough comparison would also require us to apply the APOGEE's ASPCAP pipeline to the new line list (instead of only applying The Payne to the old line list), something that we do not have the tool to perform ourselves. We will defer such detailed explorations to future studies.

%
%
%
%
%
%

\subsection{Astrophysical verification of The Payne}
\label{sec:testing-the-Payne}

In this section we present two important tests of \tP: first, we compare newly generated {\it ab initio} models that were not included in the training step to models predicted from \tP. This step directly tests interpolation errors in the training of the neural networks. Second, we fit noiseless models with \tP\, to see how well we can recover stellar labels in the case of perfect synthetic models. This step tests how much any interpolation errors in flux space translates into uncertainties in determining accurate stellar labels.

Fig.~\ref{fig4} shows how well \tP\, interpolates synthetic spectra. We trained on 2000 training spectra and test on the additional 850 validation synthetic spectra that are not used in training. The top left panel shows a small segment of wavelength range, comparing \tP\, interpolation with the {\it ab initio} calculated spectra. The upper case illustrates a spectrum where the interpolation error is small (<0.1\%). Most of the validation spectra are in this category. The lower case shows one of the few extreme cases where the interpolation is poor (>1\%).

The top right panel shows the absolute interpolation errors for different synthetic spectra at different temperature ranges, taking the median over all wavelength pixels. For each synthetic spectrum, the median interpolation error is only about 0.1\% with \tP, more accurate than the typical S/N observed by APOGEE. Cooler stars have slightly larger errors because there are more spectral features in cool stars and the imperfectness of continuum normalization becomes more severe. We note however that in some cases, the errors can be $>1\%$. We tested that including 10 times more training data and increasing (or decreasing) the number of neurons does not improve these cases. We will leave the fine-tuning of the network architecture and loss function as well as the tailoring of specific regularization to mitigates these extreme cases to future studies. Nonetheless, although not shown, we also tested that with a quadratic model, the interpolation errors are typically an order of magnitude larger, which is not surprising given the large range in $T_{\rm eff}$ and $\log g$ under consideration.

The bottom panels illustrate the pixel-by-pixel interpolation errors, averaging over validation spectra. Plotted on the bottom left panel is the median errors for a randomly selected wavelength segment. Typical pixel-by-pixel errors for \tP\, are about 0.1\%. The results over all wavelength pixels are summarized in the bottom right panel, which illustrates the cumulative number of wavelength pixels as a function of interpolation errors. The solid lines show the median as before, and the dashed lines indicate the 95 percentile (2$\sigma$) -- i.e., pixel-by-pixel, more than half of validation spectra have interpolation errors smaller than the solid line with \tP, and more than 95 percentile of the validation spectra are within the interpolation errors illustrated by the dashed line.

Having established that \tP\, can interpolate models well, we will now investigate how much the interpolation error in flux space translates into accuracy error in determining stellar labels, i.e., in the limit of perfect spectral models with no noise, how well \tP\, can recover the stellar labels. This will set a lower limit floor on how accurate (not precision) \tP\, can recover stellar labels. Fig.~\ref{fig5} shows the recovery of stellar labels of the validation spectra by fitting (noiseless) validation spectra with \tP. Throughout this study, we fit spectra by minimizing the $\chi^2$ of the interpolated model to the fitting spectra. The $\chi^2$ minimization is performed using {\sc scipy.optimize.curvefit}. When fitting real observed APOGEE spectra, we also take into account the reported uncertainties for individual pixels; pixels masked out by spectroscopic mask are set to have infinite uncertainties. We have tested that initializing at different initial points for the $\chi^2$ minimization results in the same solutions. This is not surprising because, at the APOGEE's resolution, most spectral features are resolved, and hence the degeneracy of stellar labels is not severe \citep{tin17a}. As such, we only run the optimization once for each spectrum. Since generating a spectrum to compare with the fitting spectrum requires only evaluating a function (the trained neural networks) which takes only milliseconds, the optimization typically only consumes one CPU second to fit for an APOGEE spectrum.

The top panel shows the $1\sigma$ of the label recovery. As shown in the red line, for the bulk of the APOGEE spectra which have $T_{\rm eff} \simeq 4500-5000\,$K, in the limit of perfect models, \tP\, can recover labels to an accuracy of $\simeq 0.02-0.1\,$dex for elemental abundances, $30\,$K for $T_{\rm eff}$ and $0.05\,$dex for $\log g$. Some elemental abundances have larger accuracy problems, but these are elemental abundances that have rather weak signatures and/or with strong blends. In practice, almost all the 15 elemental abundances (C, N, O, Mg, Al, Si, S, K, Ca, Ti, Cr, Mn, Fe, Ni) that we will focus in the APOGEE example study have accuracy better than $\sim 0.05\,$dex. The blue line shows the accuracy for stars cooler than $4500\,$K (e.g., Arcturus). Despite having more spectral features, the typical accuracy for cooler stars is two times worse due the larger interpolation errors, as already illustrated in Fig.~\ref{fig4}. We also note that while the there might be biases for individual stars of $0.03-0.1\,$dex, the bottom panel shows that, if the training sample is a fair representation of the global APOGEE chemical distribution, there is no strong overall biases due to the interpolation error. Plotted is the median deviation of the validation spectra fit to the assumed input. For all abundances, the overall biases is typically less than $0.01\,$dex.

Importantly, we emphasize that the results show the accuracy of \tP\, instead of precision because at a given stellar label, although \tP\, could incur a bias, the differential recovery can still be very precise. As we will see in the APOGEE example application below, we achieve an elemental abundance precision of about $0.03\,$dex for all elemental abundances by fitting the APOGEE spectra.
\begin{figure*}
\centering
\includegraphics[width=1.0\textwidth]{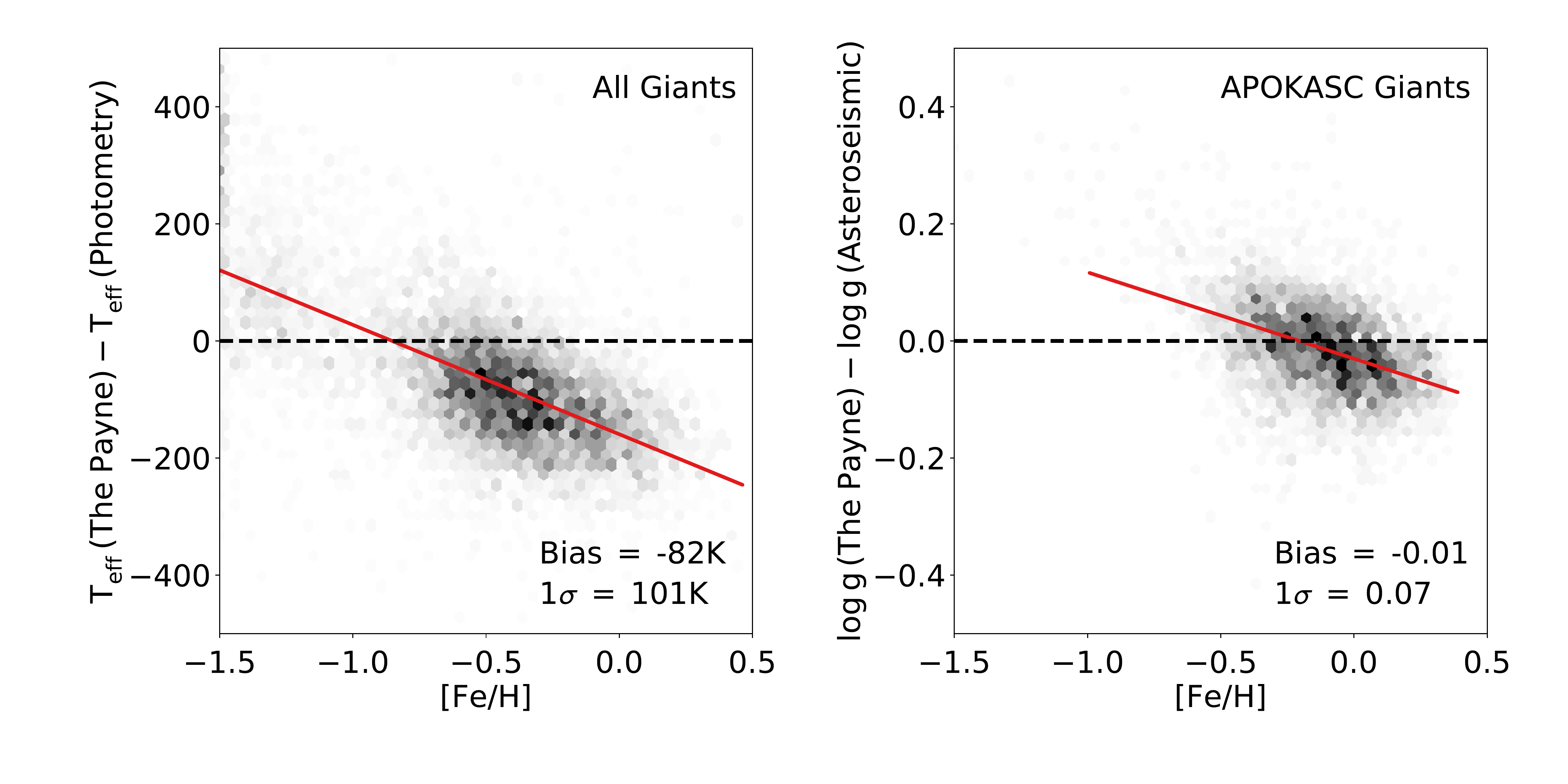}
\caption{Accuracy of \tP\, $T_{\rm eff}$ and $\log g$ estimates compared to independent external constraints. The left panel shows the differences between spectroscopic $T_{\rm eff}$ estimated by \tP\, compared to $T_{\rm eff}$ derived from 2MASS infrared photometry following \citet{gon09} (with \tP\, metallicity as input). We only consider giants that have small extinction $E(B-V) < 0.02$. galactic latitude $|b| > 30$ and $0.1 < J-K < 0.9$ to ensure the accuracy of the photometric estimations. The right panel shows the differences between spectroscopic $\log g$ from \tP\, compared to APOKASC (v3.6.5) asteroseismic $\log g$. \tP\, $T_{\rm eff}$ and $\log g$ agree with these external calibrators to about $200\,$K and 0.1 dex, and exhibit with a weak metallicity dependence. The red lines illustrates the best fit linear relation of the metallicity trend.}
\label{fig8}
\end{figure*}

%
%
%
%
%
%

\section{An illustration of The Payne: 25 Stellar labels from APOGEE data}
\label{sec:implementing-apogee}

As a specific application and illustration of \tP, we fit the entire APOGEE DR14 data set, consisting of $\sim 270$,$000$ spectra. We only consider the combined APOGEE spectra (instead of individual visits) throughout this study. We train \tP\, with only $2000$ {\it ab initio} model spectra, and then fit for 25 stellar labels. We also fit for the radial velocity at the same time during the fit to avoid any radial velocity residual from the APOGEE reduction pipeline. When comparing to APOGEE DR14 values, we will refer to the official APOGEE pipeline, ASPCAP, values, instead of the values from {\it The Cannon}.

%
%
%
%
%
%

\subsection{Fitting the Sun \& Arcturus at APOGEE resolution}

We start out by illustrating how well \tP\, does in fitting Arcturus and the Sun at the resolution of APOGEE (Fig.~\ref{fig6}). We generated 100 realizations of Arcturus' and the Sun's APOGEE spectra, just differing by Poisson noise of the spectra  (S/N$\sim 400$). The violin plots in Fig.~\ref{fig6} show the deviations of our fit of all 100 realizations from the Arcturus and Solar benchmark values adopted from \citet{ram11} and \citet{asp09}. The solid black line shows the corresponding APOGEE DR14 values. Overall, \tP\, shows comparable deviations from the benchmark values as APOGEE DR14, about $0.1\,$dex for elemental abundances. Part of the deviations is due to the interpolation accuracy error described, but they are also partially contributed by the imperfect spectral models. For individual objects, performing full spectral fitting with \tP\, can be more susceptible to model imperfection due to the covariant spectral features, especially with the lenient cut that we made which keeps almost the full APOGEE spectrum. If we were to make a more stringent cut for the spectroscopic mask, i.e. 0.5\% error instead of the fiducial 2\% error adopted, as shown in the red dashed lines, the accuracy can get better, with the exception of V which only has a very weak feature at the Solar temperature. But this comes at the expense of the precision of stellar labels for the overall sample because, as illustrated in Fig.~\ref{fig2} and Fig.~\ref{fig3}, with a more stringent cut, we discard a significant portion of the spectra. Therefore, we adopt the fiducial spectroscopic mask throughout this study.

%
%
%
%
%
%

\subsection{$T_{\rm eff}$ \& $\log g$}
\label{sec:teff_and_logg}

Fig.~\ref{fig7} shows how well \tP\, can recover stellar parameters ($T_{\rm eff}$, $\log g$, [Fe/H]) for both giants and dwarfs with a single self-consistent training model. The left panel shows the values obtained by \tP\,, and the right panel shows the APOGEE DR14 {\em calibrated} counterparts. APOGEE DR14 does not provide calibrated stellar parameters for dwarfs and sub-giants as they found that the current pipeline struggles to provide reliable estimates for non-giants \citep[e.g.,][]{hol15}. Overplotted in the both panels are the MIST isochrones, but at different stellar ages. \tP\, derives $T_{\rm eff}$ and $\log g$ that are consistent with the MIST isochrones at 7 Gyrs, and the estimates show less scatter at the metal poor end for the giants compared to APOGEE DR14. The APOGEE calibrated their values with the photometric $T_{\rm eff}$ and the asteroseismic $\log g$ as we will discuss below, and the calibrated values are more consistent with 1.5 Gyrs old MIST isochrones, which might be too young for the bulk of the APOGEE data. It thus suggests that there is a discrepancy between the photometric $T_{\rm eff}$, which the APOGEE values calibrated against, with the spectroscopic $T_{\rm eff}$ from \tP, and the MIST isochrones at the $100\,$K level. The figure shows that APOGEE DR14 calibrated values also generally favor more metal-rich estimates than \tP. But this is largely due to their calibration with photometric temperature as we will discuss below.

\tP\, does not perform as well for the cooler dwarf stars ($T_{\rm eff} < 4000\,$K) especially for metal poor stars ([Fe/H]$\,<-0.5$). This could due to multiple reasons. For example, our adopted line list is only calibrated against hotter and more metal rich stars -- Arcturus ($T_{\rm eff} \simeq 4300$) and the Sun ($T_{\rm eff} \simeq 5800$). Moving forward, spectral models built from an atomic line list that has been calibrated against a wider array of stars will be very valuable. The failure in the metal-poor dwarf regime could also be due to a breakdown of the assumptions of LTE. 

As shown in Fig~\ref{fig7}, the $T_{\rm eff} - \log g$ for dwarfs also exhibits a larger spread in the $T_{\rm eff} -\log g$ Kiel diagram than what is predicted by the stellar evolution models. Part of this larger spread could be due to the fact a non-negligible fraction of the main sequence stars could be unresolved binaries. Fitting single star models to binaries will incur biases which manifest itself as a thicker sequence in the Kiel diagram \citep{elb18a}. It is beyond the scope of this paper to fit for binaries, but we caution that the single star assumption can compromise the abundance precision that we obtain for dwarfs. For giants, the single star assumption is less a problem because the giant will outshine its dwarf companion, and giant-giant binaries are rare. We refer readers to \citet{elb18b} where \tP\, was adopted to fit for main sequence binaries by fitting a mixture of (data-driven) stellar models.
\begin{figure*}
\centering
\includegraphics[width=1.0\textwidth]{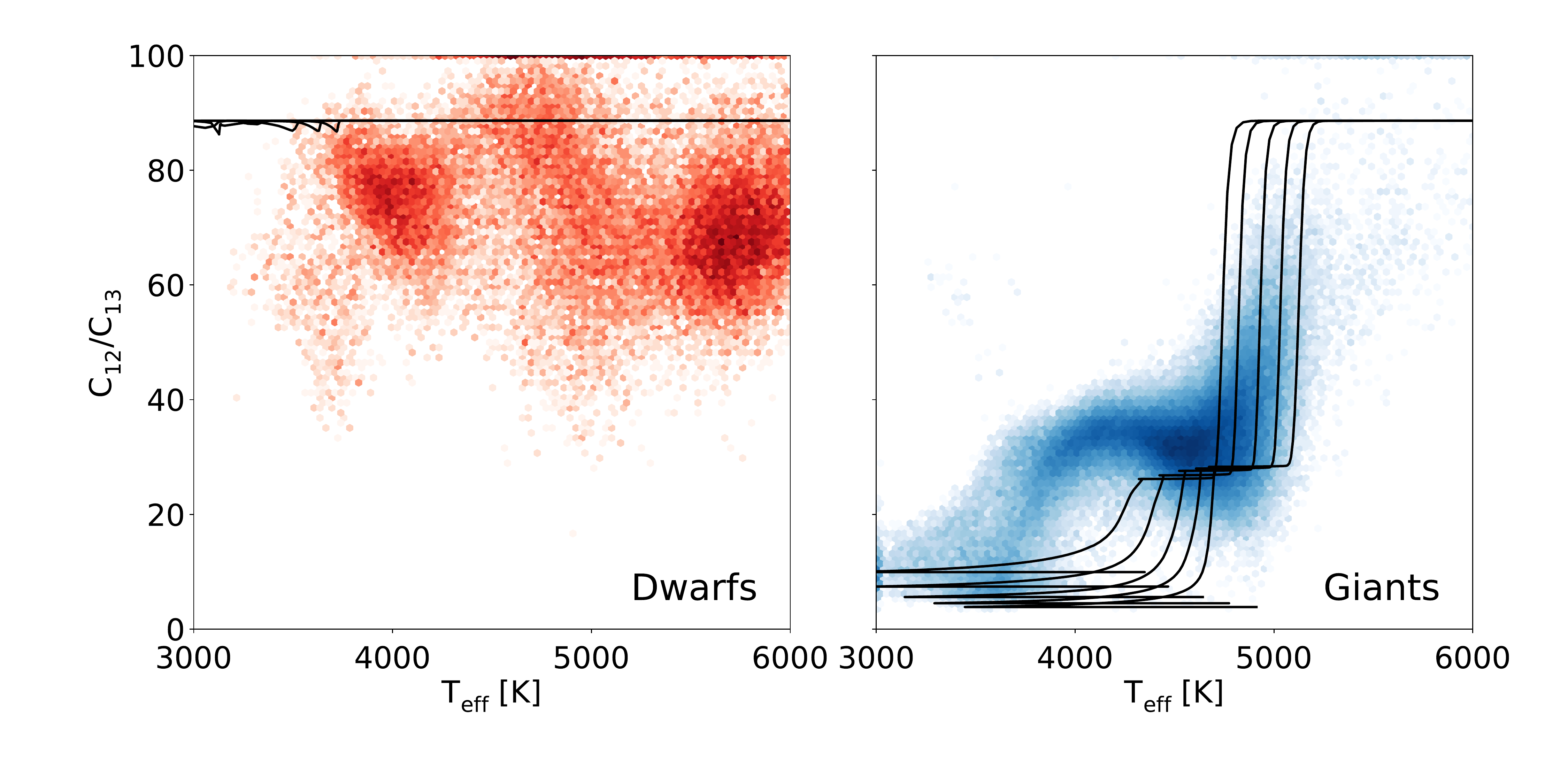}
\caption{$C_{12}/C_{13}$ estimates from \tP. The $C_{12}/C_{13}$ estimates for dwarf stars ($\log g > 4$) are shown on the left, those for giants ($\log g < 4$) on the right; we only show hexbins that have more than 10 stars. Dwarf stars typically show $C_{12}/C_{13}$ from 60 to 90, but the exact values are not well constrained by \tP\, in APOGEE because the spectra show little change for $C_{12}/C_{13} > 50$. The black solid lines reflect stellar evolution models from MIST, with a stellar age of 7 Gyrs old and metallicity $-0.5 < {\rm [Fe/H]} < 0.5$. On the left panels, we show the isochrones for the main sequence, and on the right the turn-off phase to the Helium core-burning red clump phase. For giants, the determined $C_{12}/C_{13}$ values from \tP\, roughly follow the expected trend with a first drastic transition in $C_{12}/C_{13}$ at $5000\,$K, caused by the first convective dredge-up and and second dip at a lower temperature. But we caution that these $C_{12}/C_{13}$ results are partially driven by the prior we impose in the training set (see text for details).}
\label{fig9}
\end{figure*}

In Fig.~\ref{fig8} we compare \tP\, estimates with $T_{\rm eff}$ and $\log g$ derived from other external means. In the left panel we compare the spectroscopic $T_{\rm eff}$ to the $J-K$ color-$T_{\rm eff}$ relation from \citet{gon09}. For this comparison we only consider giants that have small line of sight extinction, i.e., $E(B-V) < 0.02$ from the SFD map \citep{sch98}, avoid the galactic disk $|b| > 30^\circ$ and have color $0.1 < J-K < 0.9$ following \citet{gon09}. In the right panel we compare spectroscopic $\log g$ for a subset of 3000 stars that have APOKASC v3.6.5 asteroseismic $\log g$ values. Without calibration, the $\log g$ estimates from \tP\, agree with the asteroseismic $\log g$ value to about $0.07\,$dex with only a weak metallicity dependence. Overplotted in red line is the best fit linear regression. We do not overplot the APOGEE DR14 values because, by definition, APOGEE DR14 $\log g$ are calibrated to match the APOKASC asteroseismic $\log g$ and the photometric temperature. As shown in the left panel, spectroscopic $T_{\rm eff}$ from \tP\, however, is typically $100\,$K cooler than the photometric $T_{\rm eff}$, and shows a dependence with metallicity. It is hard to speculate what causes these trends, but it could either be inflicted by the inherent differences between H-band spectroscopic and photometric temperature, since APOGEE DR14 uncalibrated values also show similar offsets, or it could simply be due to the imperfect spectral model and line list. We found that imposing a more stringent spectroscopic mask does not resolve this issue, indicating that the cooler temperature is favored by our spectroscopic model and is not due to interpolation error. But as we will see, even without calibrating this relation, the derived stellar labels from \tP\, seem to agree well in other plausibility tests that we will present below. So we choose not to calibrate the temperature and will leave the more detailed study of this discrepancy to future studies \citep[e.g.][]{cho18}.

One particularly interesting aspect of \tP\, as shown in Fig.~\ref{fig7}, is that besides deriving stellar parameters for the dwarfs, \tP\, also yields reasonable $T_{\rm eff}$ and $\log g$ for the giants on the cooler end, around 3500$\,$K to 4000$\,$K. In fact, we found that fitting $C_{12}/C_{13}$ is crucial to get $T_{\rm eff}-\log g$ that are consistent with the isochrone at the cooler end for the giants. Since $C_{12}/C_{13}$ spectral features are highly blended with other features, $C_{12}/C_{13}$ can only be reliably derived with a full spectral fitting with all stellar labels simultaneously, an area where \tP\, excels.
\begin{figure*}
\centering
\includegraphics[width=1.0\textwidth]{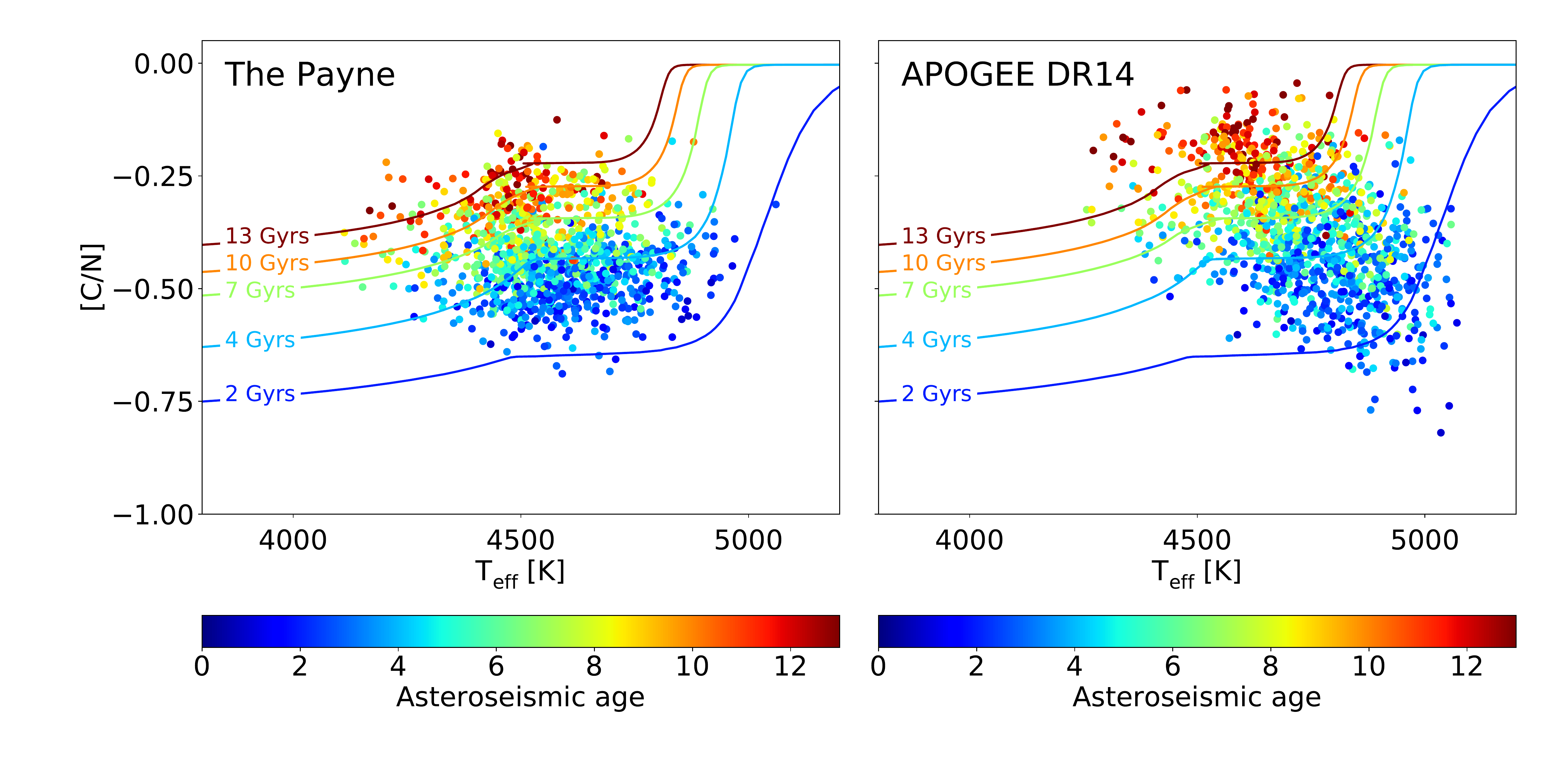}
\caption{\tP\, measures accurate and precise [C/N] ratios for field stars. We plot the spectroscopic [C/N] ratios for the subsample of $-0.1 < {\rm [Fe/H]} < 0.1$ of the APOKASC sample, color-coded with their corresponding precise asteroseismic ages. On the left panel, we show the estimates from \tP, and on the right panel the APOGEE DR14 values. The solid lines of different colors show the [C/N] predictions from various MIST isochrones of Solar metallicity. \tP\, measurements agree better with the isochrones and show a smaller scatter and bias compared to APOGEE DR14. The excellent agreement with the isochrones affirms the ability to infer stellar ages directly from abundance measurements.}
\label{fig10}
\end{figure*}

\begin{figure*}
\centering
\includegraphics[width=1.0\textwidth]{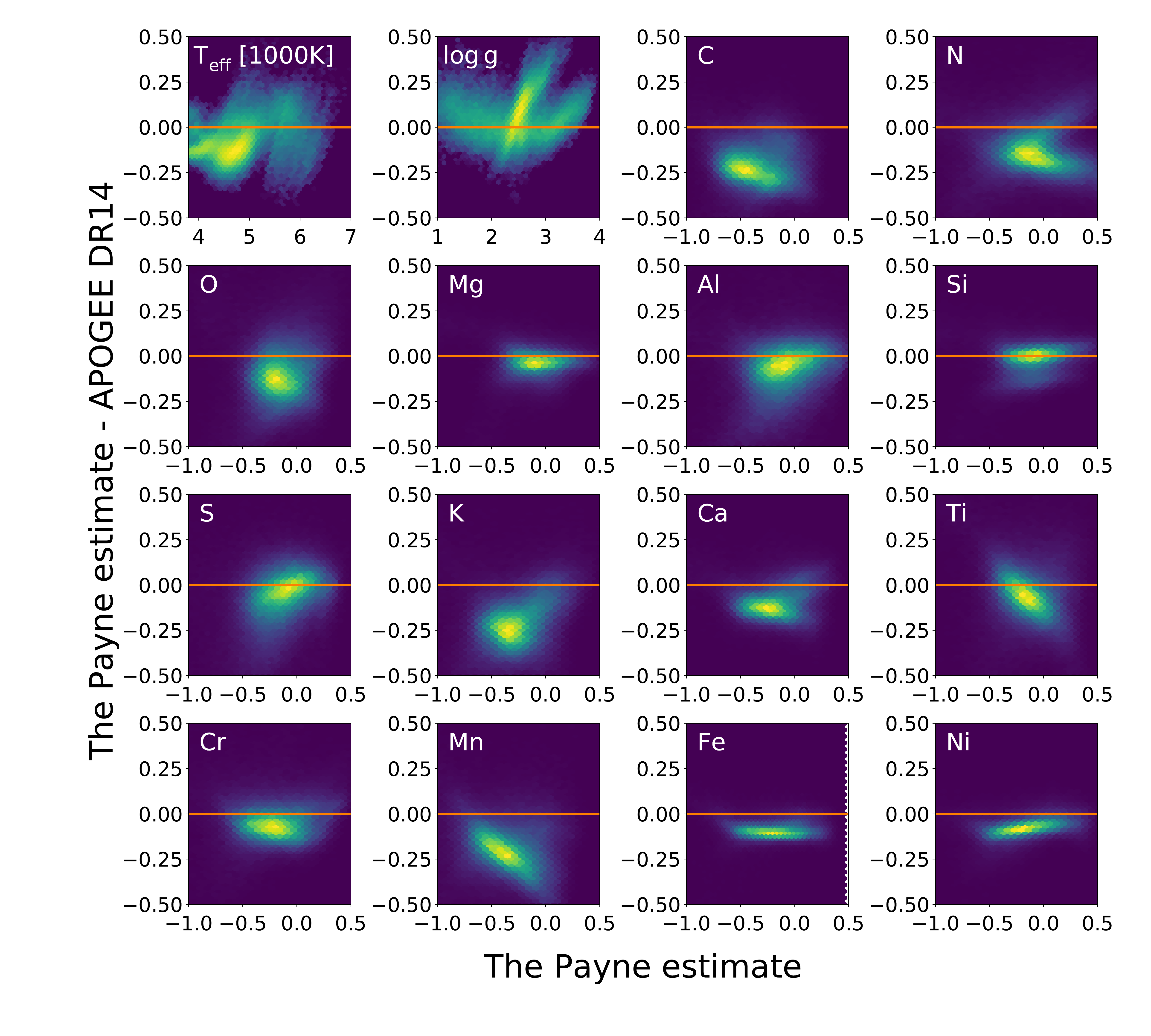}
\caption{Comparison of \tP\, estimates with the APOGEE DR14 calibrated stellar labels. We perform a full spectral fitting for all stellar labels as shown in the plot, as well as fitting $v_{\rm micro}, v_{\rm macro}$ and $C_{12} / C_{13}$ simultaneously. We do not compare Cu since APOGEE does not provide Cu abundances in DR14. Note that, for $T_{\rm eff}$ and $\log g$, we plot the density plot in log scale to emphasize the contrast since most APOGEE are concentrated near the red clump $T_{\rm eff}$ and $\log g$. In general, \tP\, prefers slightly ($\sim0.08\,$dex) more metal-poor estimates than APOGEE DR14 calibrated values. There is a visible deviation in $\log g$ around $\log g = 2.5$; \tP\, $\log g$ estimates for red clumps are slightly higher compared to the {\em calibrated} APOGEE values. }
\label{fig11}
\end{figure*}

\begin{figure*}
\centering
\includegraphics[width=1.0\textwidth]{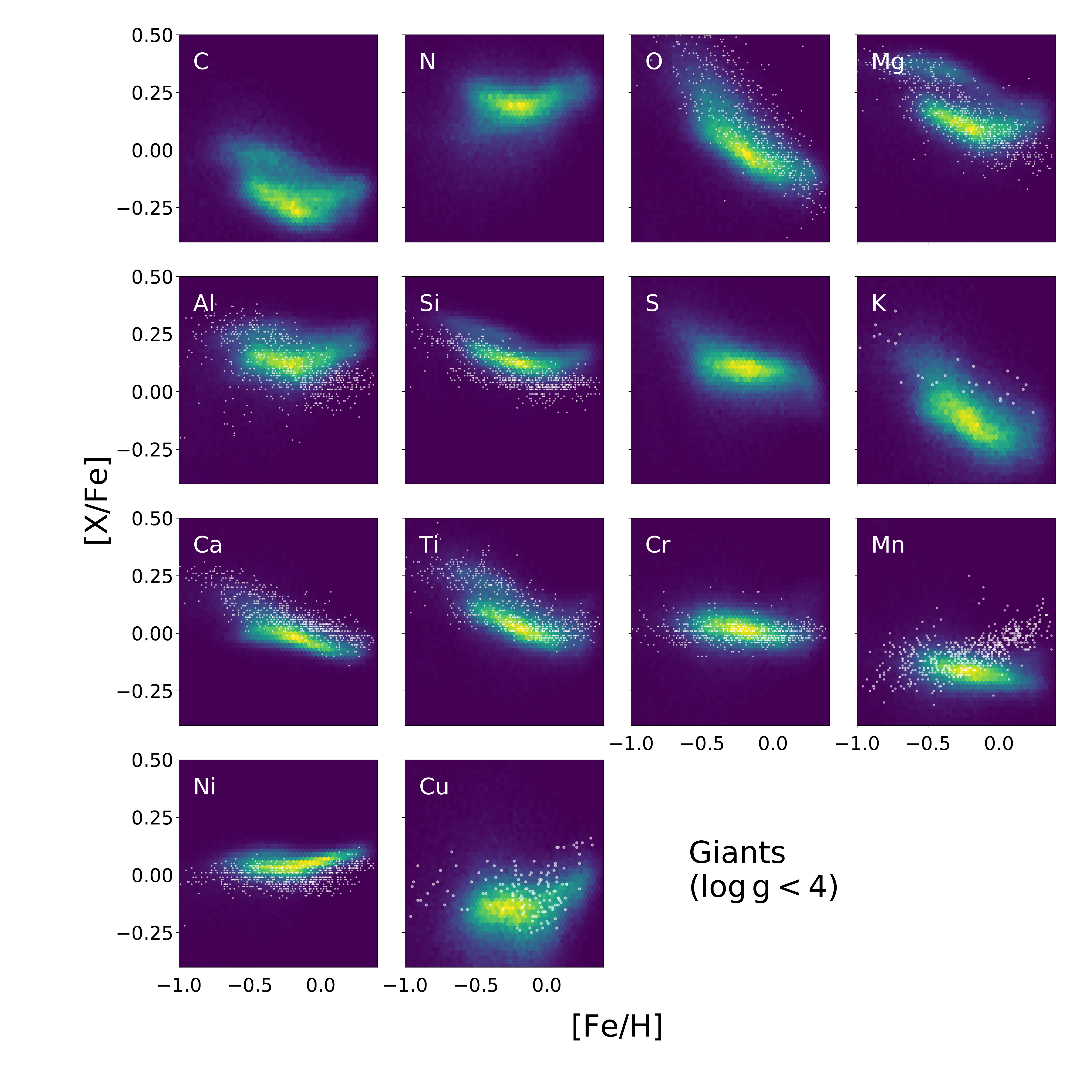}
\vspace{-0.85cm}
\caption{$[X/{\rm Fe}]-[{\rm Fe/H}]$ for 14 elements for the APOGEE DR14 giants ($\log g < 4$), derived with \tP. The background shows the density plot of the label estimates from \tP. Overplotted in white symbols are a compilation of literature values from \citet{mis11}, \citet{ben14}, \citet{nis14}, \citet{bas15}, and \citet{zha16}. \tP\, recovers the separation of the high- and low-$\alpha$ sequences. Elemental abundances from \tP\, in [X/Fe] agree well with the literature values without any calibration. Nonetheless, there is a slight offset in Si, K and Ni compared to the literature values. Also our models prefer a flat [Mn/Fe] trend which is in contrast with the literature values. Notably, \tP -derived Ti abundances follow the expected trend, overcoming a persistent issue in APOGEE DR14.}
\label{fig12}
\end{figure*}

%
%
%
%
%
%

\subsection{$C_{12}/C_{13}$ \& C/N}
\label{sec:c12c13_and_cn}

The flux variation dependence on $C_{12}/C_{13}$ is a particularly difficult to model. As already shown in Fig.~\ref{fig1}, the flux variation as a function $C_{12}/C_{13}$ has a sharp transition. Above $C_{12}/C_{13} \simeq 50$, the spectral dependence is very weak, and below $\sim50$ the flux varies strongly with $C_{12}/C_{13}$. Since carbon molecular features contribute significantly to the H-band APOGEE spectra, $C_{12}/C_{13}$ alters the spectra in a significant way. On the one hand, it implies that fitting $C_{12}/C_{13}$ is not only astrophysical interesting, it can also be crucial as part of the spectral fitting, without which the stellar parameters may be biased. But on the other hand, we found that, in the limit of imperfect models, if we do not impose a prior $C_{12}/C_{13}$, the $C_{12}/C_{13}$ features can be wrongly adopted to adjust the global fit to get a lower $\chi^2$. Therefore, as discussed in Section~\ref{sec:priors}, we assume a weak prior for $C_{12}/C_{13}$ from stellar evolution models.

Fig.~\ref{fig9} shows the $C_{12}/C_{13}$ values estimated with \tP\, for all APOGEE stars. On the left, we show the $C_{12}/C_{13}$ values for dwarfs (with $\log g > 4$), and on the right for giants ($\log g < 4$). Overplotted in black lines are the MIST isochrones for the respective evolutionary states, assuming a stellar age of 7 Gyrs old, and metallicity [$Z$/H] ranging from -0.5 to 0.5. The $C_{12}/C_{13}$ values for dwarfs are less well constrained and have a larger scatter from the MIST prediction because the spectral response with respect to $C_{12}/C_{13}$ at $C_{12}/C_{13} > 50$ is weak and yields almost identical spectrum (see Fig.~\ref{fig1}). As for the giants, the $C_{12}/C_{13}$ values roughly agree with the MIST isochrones, with a sharp transition around 5000$\,$K due the the first convective dredge-up and follow by a second transition as the stars ascend in the HR diagram in the red-giant branch and reach a lower temperature. But the transition temperature seem to be smaller than the predictions from stellar evolution models.

We caution readers not to over interpret the $C_{12}/C_{13}$ results as we have assumed a prior for the $C_{12}/C_{13}$ in the training set. One of the current challenges of full spectral fitting is that, in the limit of imperfect models, one stellar label, such as $C_{12}/C_{13}$, may in effect "do the work" of another stellar label. As discussed, the reason to include $C_{12}/C_{13}$ is merely to ensure that the stellar parameters are robust at the cooler giant end since it contributes significantly at the cooler end due to the strong features as well as the second dredge-up. It also shows that $C_{12}/C_{13}$, in principle, can be fitted simultaneously with all other labels employing \tP.

Besides $C_{12}/C_{13}$, the [C/N] ratio of stars will also be modified due to convective dredge up during the giant phase. In fact, the [C/N] ratio has been shown to excellent stellar mass indicators for giants \citep{mar15,nes16,ho17b}; how much the dredge-up affects the [C/N] ratio depends crucially on the stellar mass. Since there is a tight correlation between stellar mass and stellar age (given a fixed metallicity), determining accurate [C/N] ratios from large spectroscopic surveys is particularly important because they are excellent age indicators for stars. In Fig.~\ref{fig10}, we overplot the [C/N] ratios of the APOKASC sample, color-coded with their corresponding asteroseismic ages, with the predictions from the MIST isochrones. Since stellar evolution predictions depend on metallicity, we restrict the APOKASC sample with $-0.1 < {\rm [Fe/H}] < 0.1$ and assume Solar abundances for the isochrones. On the left-hand side, we show the results from \tP, and APOGEE DR14 on the right-hand side. \tP\, values agree better with the isochrones and show a reduced scatter and bias, especially for the older stars, indicating that our C to N abundances are likely more accurate. The excellent agreement between the stellar evolution models with spectroscopic indices also demonstrates that by fitting all stellar labels self-consistently and simultaneously, the improved spectral models and stellar evolution modes can be accurate enough to allow for a direct inference of stellar ages from spectroscopic indices, going beyond data-driven models.

\begin{figure*}
\centering
\includegraphics[width=1.0\textwidth]{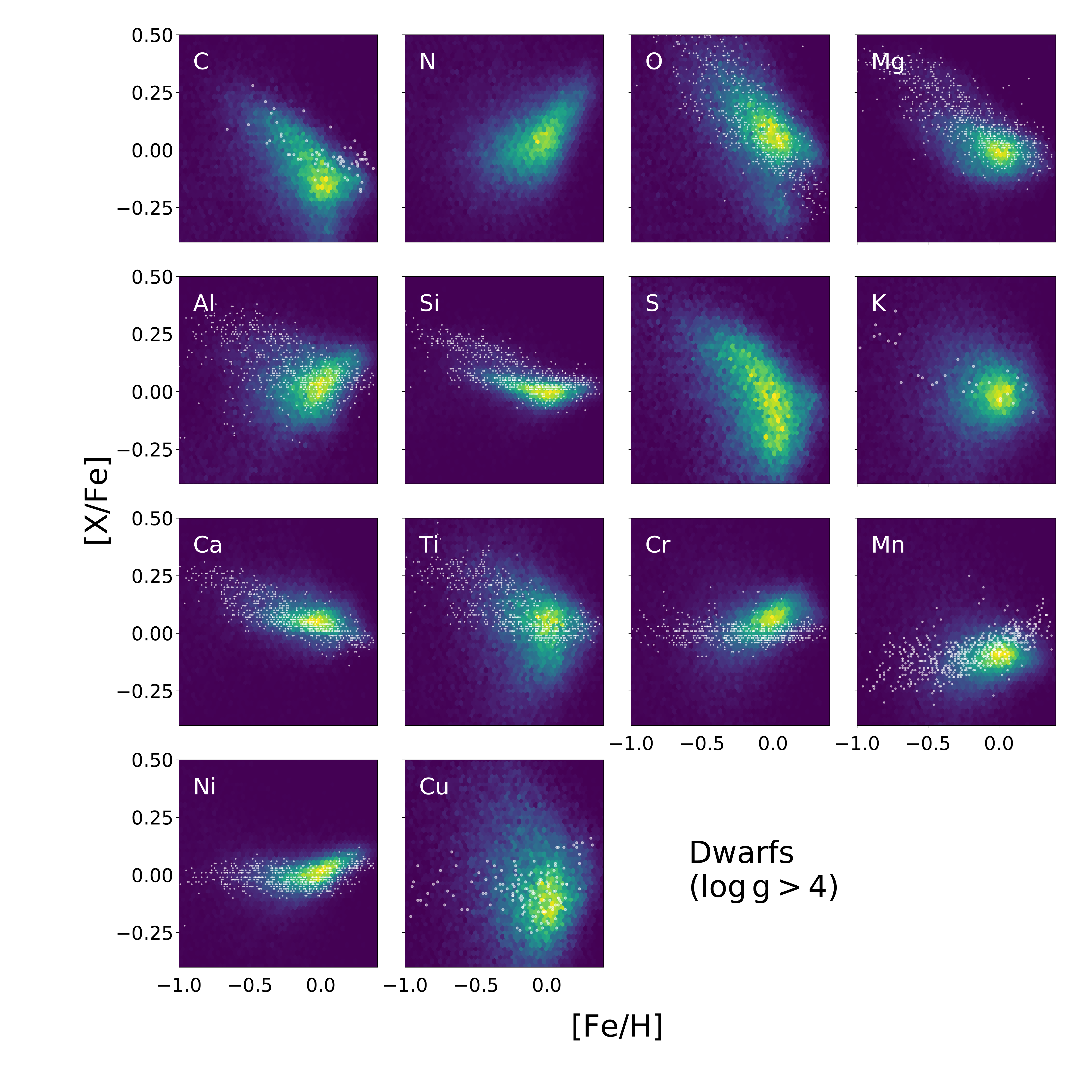}
\vspace{-0.85cm}
\caption{The same as Fig.~\ref{fig12}, but for dwarf stars. The dwarf abundances estimated by \tP\, are consistent with the giant abundances with a few exceptions. The C and N abundances for the dwarfs are expected to differ from the giants due to convective dredge-up. The Al, Si, K, Mn and Ni abundances for the dwarfs agree better with the literature values, suggesting that the discrepancies seen in the giants are mostly spectral model related. On the other hand, the distribution of Cr abundances among the dwarfs favors an upward trend, and the Ti and S distributions have a larger spread than among the giants. The precision for the dwarfs can also be affected by unresolved binaries, which could constitute a large fraction of the dwarf stars in APOGEE, and might explain the marginally larger abundance spread for the dwarfs. Most dwarf stars are in the low-$\alpha$ sequence because they are, on average, closer than the giants.}
\label{fig13}
\end{figure*}

%
%
%
%
%
%

\subsection{Element abundance patterns}

Elemental abundances are often derived from individual spectral lines, one element at a time. A key goal of \tP\, is to demonstrate that all elemental abundances can be measured from stellar spectra directly from a simple $\chi^2$ fit by fitting all elemental abundances and stellar parameters simultaneously. In this study, we fit for 20 elemental abundances, namely C, N, O, Na, Mg, Al, Si, P, S, K, Ca, Ti, V, Cr, Mn, Fe, Co, Ni, Cu, Ge, all elemental abundances show visible absorption lines from our line list in the H-band. As already shown in Fig,~\ref{fig5}, in the limit of perfect models and data, all of these elemental abundances can be extracted with \tP.

However, we found that 5 elemental abundances (Na, P, V, Co, Ge) cannot be reliably derived with the current implementation of \tP, an issue also well-diagnosed in APOGEE DR14 \citep[e.g.,][]{hol15}. These elements exhibit large scatter in an $[X/{\rm Fe}]-[{\rm Fe/H}]$ diagnostic plot or a large scatter in the precision test (Section~\ref{sec:abundance-precision}). Elements like Na, P, V have only weak features ($<1\%$ change in flux for $\Delta [X/{\rm H}] = 0.05$) in the H-band, and unfortunately, the features are also often blended with the telluric sky lines, an issue compounded by the current interpolation errors from \tP. Although we derive estimates from these elemental abundances, we decided that they are not to be trusted. The reason for a large spread in Co, Ge in an $[X/{\rm Fe}]-[{\rm Fe/H}]$ diagnostic is unclear because each of these elements does have a single strong feature in H-band, similar to K, and we have no problem getting reasonable K measurements as shown below. We will defer a more detailed study of the problems to a forthcoming paper. We will focus on the remaining 15 elements, 14 of which (except for Cu) have been reliably determined in APOGEE DR14 for comparison and only consider stars with a fitting reduced $\chi_R^2 < 50$.

Fig.~\ref{fig11} shows the comparison of \tP\, estimates with the calibrated values from APOGEE DR14, showing a generally good agreement to the level of $100\,$K in $T_{\rm eff}$, $0.1\,$dex in $\log g$ and $0.1\,$dex in $[X/{\rm H}]$. \tP\, favors slightly metal-poor estimates, as already discussed in light of Fig.~\ref{fig8}. \tP\, spectroscopic estimates prefers lower temperatures, compared to the APOGEE DR14 values that are calibrated to photometric temperatures. As [Fe/H] and $T_{\rm eff}$ estimates are covariant \citep[e.g.,][]{tin17a}, this leads to more metal-rich estimates for elemental abundances. Another noticeable deviation is around $\log g \simeq 2.5$. Also shown in Fig.~\ref{fig7}, the $\log g$ values for red clump stars from \tP\, are overestimated compared to stellar evolution models. This discrepancy is also consistent with APOGEE uncalibrated values. The reason for this discrepancy is unknown; one possibility is the lack of fitting the helium abundance. It is conceivable that helium abundance differences between the RGB stars and the red clump stars could explain the $\log g$ discrepancy \citep[e.g.,][]{yu18}.

Fig.~\ref{fig12} shows the $[X/{\rm Fe}] - [{\rm Fe/H}]$ derived with \tP. The background demonstrates the elemental abundances estimated by \tP\, of the giant stars ($\log g < 4$). Overplotted in white symbols are the literature values. We consider \citet{ben14} to be the main reference literature which provides, in this plot, abundances for O, Na, Mg, Al, Si, Ca, Ti, Cr and Ni. This main sample is complemented by C abundances from \citet{nis14}, K abundances from \citet{zha16}, Mn abundances from \citet{bas15}, Cu abundances from \citet{mis11}. For [Fe/H], we adopt [Fe/H] from the same catalog to avoid systematics across different surveys. \tP\, attains reasonable $[X/{\rm Fe}]-[{\rm Fe/H]}$ without any external calibration. The separation of the high-$\alpha$ versus the low-$\alpha$ sequence is clearly visible across all $\alpha$-elements. Notably, we attain a Ti trend that is consistent with the literature values -- resolving one of the persistent problems in APOGEE \citep[e.g.,][]{hol15}. There is a $0.1\,$dex discrepancy between the literature values and \tP\, estimates for Si, K, and Ni. But we note that the K abundances from \citet{zha16} adopts NLTE models. \tP\, also favors a flat [Mn/Fe] trend, which is at odd with the literature [Mn/Fe] trend.

One important improvement coming from \tP, as already demonstrated in Fig.~\ref{fig7}, is the determination of stellar labels for APOGEE dwarf stars. Fig.~\ref{fig12} and Fig.~\ref{fig13} demonstrate that \tP\, recovers consistent abundances for both dwarfs and giants with a few key differences. First, the carbon abundances for the dwarfs are higher than the giants, and at the same time, the nitrogen abundances are lower as expected due to convective dredge up. Second, since dwarf stars are dimmer, they are dominated by stars that are closer to the Sun, and hence the dwarfs show a more prominent low-$\alpha$ sequence and have relatively fewer high-$\alpha$ stars. The dwarf abundances also seem to agree better with the literature values for Al, Si, K, Mn and Ni. Since most of the literature values are derived from main sequence dwarf stars, this agreement is encouraging and might suggest that the discrepancies between Fig,~\ref{fig12} and Fig.~\ref{fig13} might partially due to the imperfect spectral models, or could also be astrophysical related, such as atomic diffusion in dwarf stars \citep{dot17}. Interestingly, \tP\, produces upward trends for both Cr and Mn for dwarfs, and thus the dwarf Mn abundances agree with the literature values but the Cr abundances do not. Disentangling the discrepancies in Cr and Mn requires an careful investigation of the line list which we will postpone to future studies.
 
Finally, the dwarf abundances as illustrated in Fig.~\ref{fig13} show a marginally larger spread than the giants, suggesting that the precision of the dwarfs stars might be inferior than the giants stars. This might not be surprising, as a large fraction of main sequence stars could be unresolved binaries. Fitting a single star model to binaries can affect the precision \citep{elb18a}. Finally, in a companion paper, \citet{elb18b} adopted the dwarf abundances in this study to build a data-driven model and successfully fit for the unresolved binaries spectra, indirectly verifying that the dwarf stellar parameters and metallicities in this study are internally consistent and robust. 
 
 \vspace{0.5cm}

%
%
%
%
%
%

\subsection{Testing \tP\, with open and globular cluster data}

In this section we explore the stellar labels derived with \tP\, for stars in open and globular clusters with APOGEE spectra. These stars serve as strong tests of \tP\, owing both to extensive literature data and also to the fact that open clusters are believed to be at least approximately chemically homogeneous. The latter fact allows us to empirically test the measurement precision of \tP\, and also to test for any systematic behaviors in the derived labels as a function of e.g., $T_{\rm eff}$.

\begin{figure}
\centering
\includegraphics[width=0.5\textwidth]{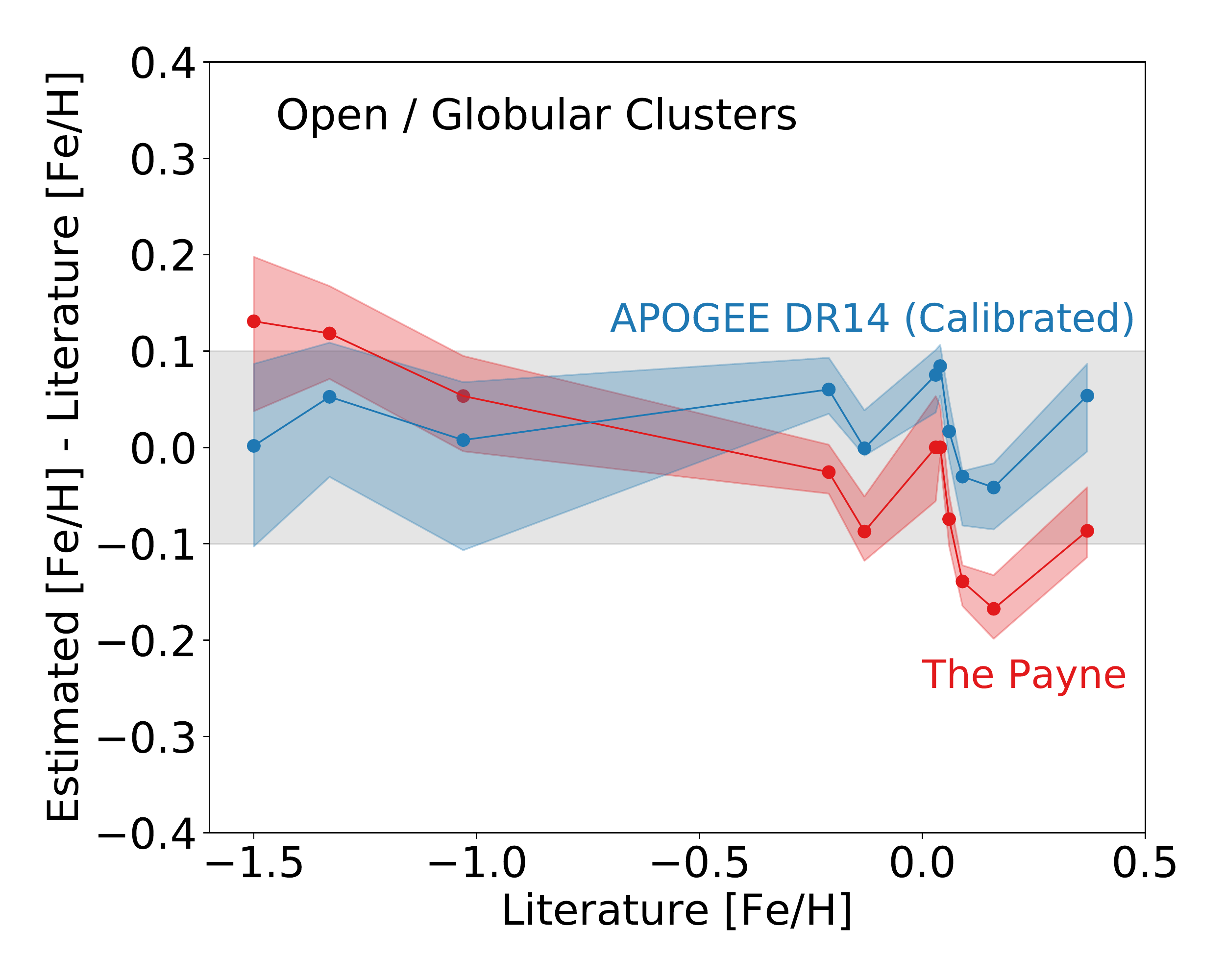}
\caption{Comparison between literature, \tP, and APOGEE DR14 [Fe/H] abundances for open and globular clusters. We compare \tP\, [Fe/H] estimates to the literature values of 11 known clusters in APOGEE (in ascending metallicity order, in square brackets are the numbers of cluster members in APOGEE: M3 [73], M5 [103], M107 [18], NGC 2158 [10], NGC 2420 [9], Pleiades [21], NGC 188 [5], M67 [24], NGC 7789 [5], NGC 6819 [30], NGC 6791 [23]) that have [Fe/H]$\, > -1.5$, the metallicity where our training set truncates. The solid lines show the median metallicity deviation, and the shaded show the $1\sigma$ range from all cluster members. Without any calibration, in the range of $-1 < \,$[Fe/H]$\, < 0$ where most of the APOGEE data resides, \tP\, derives accurate metallicities that are consistent with the literature values to $0.05\,$dex. There is a bias of $0.1\,$dex for the more metal-rich and metal-poor ends. APOGEE DR14 does not show any global trend because the calibrated values from APOGEE plotted here are calibrated against these clusters.}
\label{fig14}
\end{figure}

\begin{figure*}
\centering
\includegraphics[width=1.0\textwidth]{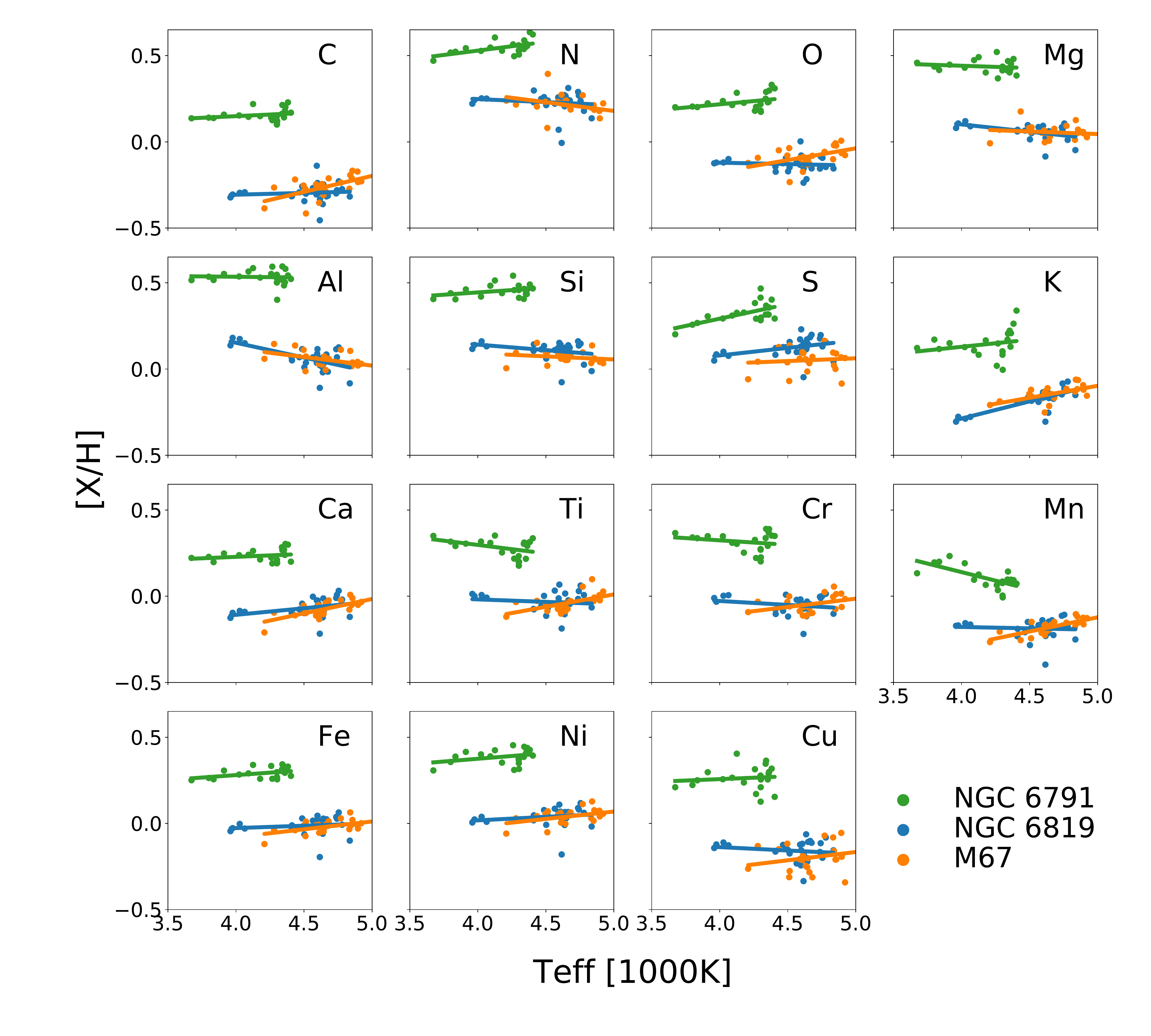}
\caption{\tP\, derives elemental abundances without any significant $T_{\rm eff}$ dependence. Elemental abundances of the members of three open clusters NGC 6819, M67 and NGC 6791 are plotted against their $T_{\rm eff}$. We also overplot the best fit linear regression in every case, merely to guide the eye. Accurate elemental abundances should show no $T_{\rm eff}$ dependence: for the vast majority of these cluster members, there is no systematic trend of abundances with $T_{\rm eff}$. \tP\, estimates do not require external abundance-$T_{\rm eff}$ calibrations within this temperature range.}
\label{fig15}
\end{figure*}

%
%
%
%
%
%

\subsubsection{The metallicity scale}

In Fig.~\ref{fig14}, we compare [Fe/H] from \tP\, with the literature values for 11 known clusters (open clusters and globular clusters) with more than 5 identified members in APOGEE and with metallicity [Fe/H]$\, > -1.5$, where our training set truncates. The open cluster members in APOGEE are identified in \citet{mes13b}. We adopt the median of all members of individual clusters to be the estimate of the cluster metallicity, and the shaded regions show the 1$\sigma$ metallicity range of all cluster members. Plotted are the differences of \tP\, and the APOGEE calibrated metallicity estimates compared to the literature values. By definition, the APOGEE metallicities show no global trend because they are calibrated against these literature values. The deviations of estimates from \tP\, shows a weak dependence with metallicity. The trend is similar to the APOGEE metallicity deviations before calibration. In fact, this behavior is likely traced back to the $T_{\rm eff}$-metallicity biases that we see in Fig.~\ref{fig8}. As the origin of these discrepancies is unclear, we choose not to calibrate our $T_{\rm eff}$ to the APOGEE scale. While we do not conform to the standards, as we have discussed in Section~\ref{sec:teff_and_logg} and in various accuracy tests throughout the paper, the APOGEE-Payne scale seems to be more consistent  with the MIST isochrone models.

Interestingly, going beyond the global trend, the APOGEE estimates and \tP\, estimates show similar relative offsets across various clusters. Since APOGEE and this study adopt very different methods (including different line lists), this suggests that the local correlated deviations from the literature values may be due to the difference between optical spectroscopy (literature values) and H-band spectroscopy (APOGEE spectra). Finally, while there is a discrepancy in metallicity, since this is temperature related, as shown in Fig.~\ref{fig12} and \ref{fig13}, it does not affect the study of $[X/{\rm Fe}]$ since the differences in the two abundances roughly cancels out as they are both caused but the differences in $T_{\rm eff}$.

\begin{figure*}
\centering
\includegraphics[width=1.0\textwidth]{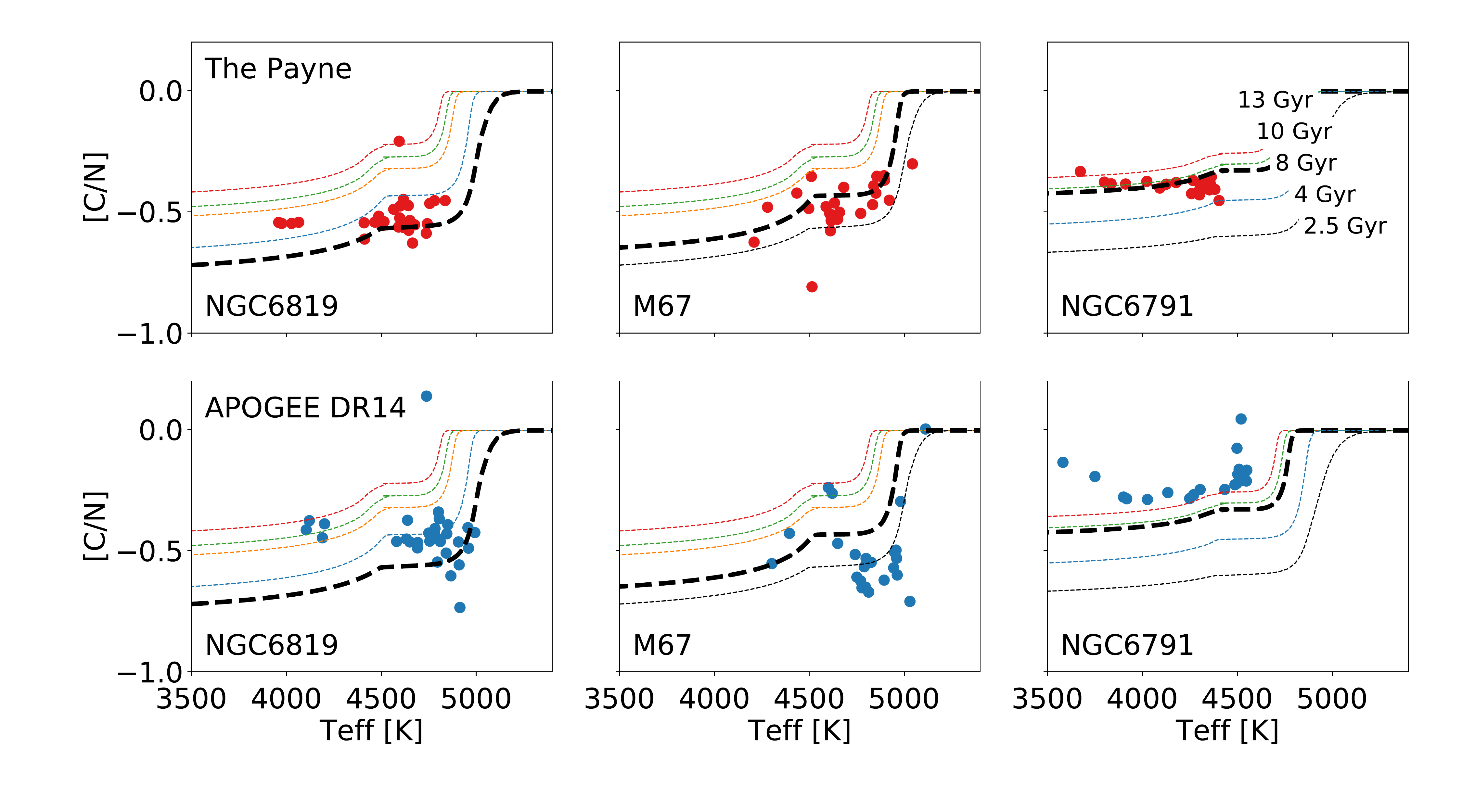}
\caption{\tP\, derives [C/N] abundance ratios that are astrophysically consistent with open cluster ages. We consider three open clusters with different stellar ages: NGC 6819 ($2.5\,$Gyrs), M67 ($4\,$Gyrs) and NGC 6791 ($8\,$Gyrs). Overplotted are the MIST stellar evolution models for different stellar ages. When evaluating the stellar evolution models, we also take into account the metallicities of these clusters: $[Z/{\rm H}] =0$ for NGC 6819 and M67, and $[Z/{\rm H}] =0.25$ for NGC 6791. The thick black dashed line in each panel shows the expected C/N ratio as a function of $T_{\rm eff}$ at these cluster ages and metallicities. In the ideal case, the data should follow a tight 1D sequence predicted by the models. The [C/N] ratio changes before and as the stars evolve up the giant branch due to convective dredge-up. The top panels show the [C/N] estimates from \tP, and the lower panels demonstrate the calibrated values from APOGEE DR14. Without any calibration, the [C/N] ratios of \tP\, agree with the MIST stellar evolution model very well suggesting that \tP\, [C/N] estimates are not only precise, they are also accurate.}
\label{fig16}
\end{figure*}

%
%
%
%
%
%

\subsubsection{Testing the abundances}

In Fig.~\ref{fig15}, we show the $[X/{\rm H}] - T_{\rm eff}$ trend of three largest open clusters in APOGEE. Open clusters are found to be very chemical homogeneous \citep{bov16,nes17}. Therefore, apart from secondary effects like dredge-up and atomic diffusion \citep{dot17}, their chemical abundances should be independent of their evolutionary state, and hence, $T_{\rm eff}$. This property is usually used to calibrate out any systematic behavior of $[X/{\rm H}]$ with $T_{\rm eff}$. As shown in Fig.~\ref{fig15}, \tP\, estimates have no significant $[X/{\rm H}] - T_{\rm eff}$ trend for both clusters, showing that our abundances display no strong systematic error as a function of $T_{\rm eff}$. However, we caution 95\% of the members from these three clusters are giants. More follow up studies of dwarf stars in these open clusters are therefore needed to test the stellar labels in the dwarf regime.

Furthermore, as discussed Section~\ref{sec:c12c13_and_cn}, the C and N abundances of stars are sensitive to stellar ages. Since open clusters have well established ages, they can also be used to check the accuracy of our C to N abundances. In Fig.~\ref{fig16}, we show the [C/N] ratios of the same three open clusters: NGC 6819 \citep[$2.5\,$Gyr, e.g.,][]{kal01,ant14}, M67 \citep[$4\,$Gyr, e.g.,][]{ric98,sar09} and NGC 6791 \citep[$8\,$Gyr, e.g.,][]{gru08}.The top panels in Fig.~\ref{fig16} show the measurements from \tP, and the bottom panels show the calibrated abundances from APOGEE DR14. Overplotted are the predictions from the MIST isochrones, taking into account the metallicities of each cluster -- $[Z/{\rm H}] = 0$ for NGC 6819 and M67; $[Z/{\rm H}] = 0.25$ for NGC 6791. The thick black dashed line in each panel shows the MIST prediction for individual clusters given their corresponding stellar ages. As shown, \tP\, [C/N] ratios agree better with the isochrones, and there is less spread indicating that our C to N abundances are likely more accurate.
\begin{figure*}
\centering
\vspace{-0.3cm}
\includegraphics[width=1.0\textwidth]{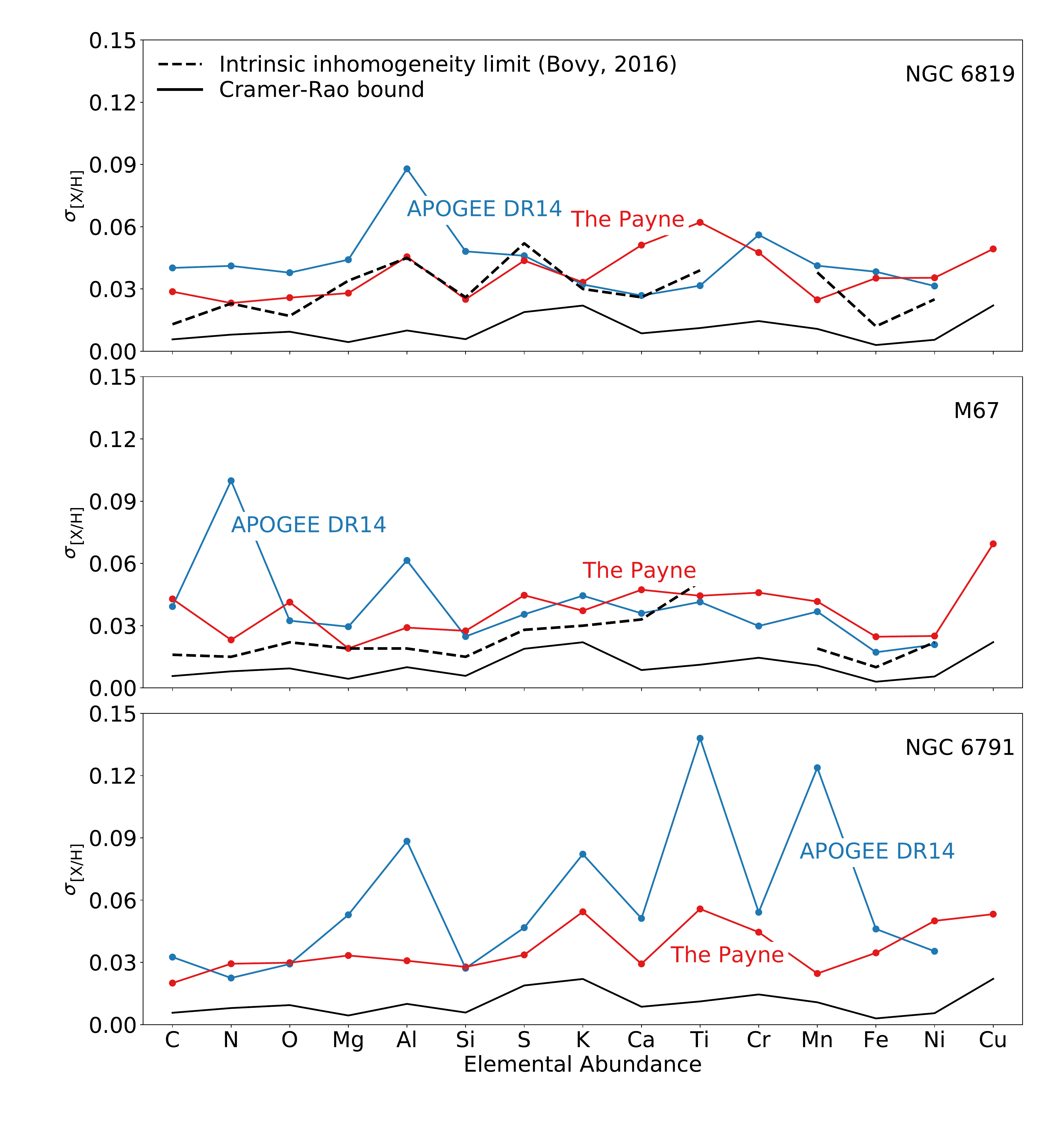}
\vspace{-0.9cm}
\caption{Empirical estimate of element abundance precision assuming intrinsically chemically homogeneous star clusters. To have a more representative of the global sample precision, we only consider cluster members that have median S/N=$\;100-300$. About 80\% of the APOGEE sample has S/N$\; > 100$. \tP\, delivers abundance precision at the $\simeq 0.03$ dex level, which is comparable to or in some cases more precise than the APOGEE DR14 precision. The three panels show the variance of various elemental abundance estimates, $\sigma_{[X/{\rm H}]}$, among members of three open clusters: NGC 6819, M67 and NGC 6791. The black dashed lines indicate the upper limits on the chemical inhomogeneity of these clusters, inferred in a data-driven approach by \citet{bov16}. The black solid lines show the Cramer-Rao bound of elemental abundances for a typical K-giant, illustrating the theoretical limit on the precision for APOGEE spectra with SNR$\,=200$.}
\label{fig17}
\end{figure*}

%
%
%
%
%
%

\subsubsection{Abundance precision}
\label{sec:abundance-precision}

Fig.~\ref{fig17} shows the elemental abundance dispersion of three open clusters discussed in the previous section. Since open clusters are chemically homogeneous, at least to the level of $0.03\,$dex \citep{bov16,liu16}, their elemental abundance dispersion gives an independent estimate of the measurement precision. Fig.~\ref{fig17} demonstrates that \tP\, obtains a precision of $0.03\,$ dex for almost all elemental abundances, more precise than APOGEE DR14 calibrated values, especially in the metal rich end (NGC 6791). We caution however that the precision achieves for individual stars clearly depend on the stellar parameters of the stars. The open clusters only probe precision at the metal-rich end. Interestingly, we found that fitting $C_{12}/C_{13}$ is the key to get more precise abundances at the metal rich end, presumably also due to a higher contribution from $C_{12}/C_{13}$, especially for the members of NGC 6791 that are, on average, cooler than the other two clusters (see Fig.~\ref{fig15}). This might be the reason why APOGEE DR14 is performing somewhat worse in precision at the metal-rich end. 

In this cluster precision test, we only consider cluster members that have median S/N$\;=100-300$, the typical S/N of the global APOGEE sample. About $80\%$ of the APOGEE sample has S/N$\;>100$. The black solid line shows the Cramer-Rao bound for a typical APOGEE K-giant with ${\rm S/N}_{\rm pix} = 200$, i.e., the best precision one could in principle achieve if there is no systematics from spectral models and interpolation \citep[see][for a detail discussion on the Cramer Rao bounds]{tin17a}. When calculating the Cramer-Rao bounds, we assume the APOGEE LSF as well as the same spectroscopic mask that we impose on the real data. We caution that while this should mimic the instrumental effect and bad telluric regions, there might be other minor instrumental/observation effects that are not being accounted in the Cramer-Rao bound. \tP\, allows us to get closer to the Cramer-Rao bound, but we are not yet reaching this fundamental limit.

We also tested how our achieved precision varies as a function of S/N by adding noise to the cluster member spectra, and found that the achieved precision is not very sensitive to the S/N. The precision consistently hovers around 0.03-0.05 for S/N$\; > 50$, and only grow as (S/N)$^{-1}$ at S/N$\; < 50$. Almost all APOGEE spectra have S/N$\; >50$. This is also consistent with both the theoretical expectation and previous empirical studies \citep{nes16, cas16, tin17a} which have demonstrated that spectra are generally information rich. Even at S/N$\sim 50$, through full spectral fitting, precise abundances can be readily achieved. 

However, why there is a precision ceiling of $\sim 0.03$ dex at higher S/N is unclear. This result is in line with previous studies \citep{bov16,liu16,nes17} illustrating that open clusters are indeed chemically homogeneous to the level of at least $0.03\,$dex. The limits derived in \cite{bov16} are plotted in black dashed line as a reference. These previous studies arrive at this conclusion either from employing a statistical argument \citep{bov16}, a data-driven approach \citep{nes17} or a more careful line-by-line differential analysis \citep{liu16}, while our result is based on direct full spectral fitting of physical spectral models to the data. It is interesting that we are not attaining the Cramer-Rao bound. Some argue that open clusters have intrinsic chemical spreads \citep{liu16} and are inhomogeneous at this level. This might well be the reason we are not reaching the best limit. But we also note that due to spectral model and interpolation imperfections, it is possible that the spread we are measuring is due to systematic errors. A further improvement of \tP\, will hopefully shed more light on the chemical inhomogeneity of open clusters.

%
%
%
%
%
%

\vspace{0.2cm}
\subsection{A catalog of stellar labels for APOGEE DR14 stars from The Payne}

We present all stellar labels ($T_{\rm eff}$, $\log g$, $v_{\rm micro}$, $v_{\rm macro}$, $C_{12}/C_{13}$ and 15 elemental abundances) in this study in an electronic form with this paper. The catalog is summarized in Table~\ref{table2}. We remove duplicated stars in the APOGEE DR14 catalog and exclude stars that have determined stellar labels that are close to the $T_{\rm eff}$, $\log g$ or [Fe/H] boundaries of our training set; we only present stars that have $3050\,$K$\, < T_{\rm eff} < 7950\,$K, $0 < \log g < 5$ and $-1.45 < [{\rm Fe/H}] < 0.45$. We also further exclude dwarf stars that have $T_{\rm eff} < 4000\,$K because as shown in Fig.~\ref{fig7}, our current models cannot determine stellar labels reliably for dwarf stars cooler than this temperature. This leaves a total of 222$,$707 stars in our catalog. 

We caution that in this catalog we keep stars that have large $\chi^2_R$ in the fitting for completeness, but we recommend readers to only use stars that show ``good'' in the "quality\_flag" column. This flag excludes all stars with $\chi^2_R > 50$, a fiducial cut we adopt in this study. It also excludes fast rotators with $v_{\rm macro} > 20\,$km/s (mostly hot stars with $T_{\rm eff} > 6000\,$K). We found that some rapidly rotating stars yield unreliable abundance patterns. But this is expected because here we do not properly account for stellar rotation $v \sin i$ and our training grid truncates at $v_{\rm macro} = 30\,$km/s, a broadening that is still too small for typical fast rotators. We will explore the inclusion of rapid stellar rotation in the future.
\begin{table*}
\begin{center}
\caption{APOGEE-{\it Payne} catalog: APOGEE stellar labels determined with \tP.\label{table2}}
\begin{tabular}{lcccccccccc}
\tableline \tableline
\\[-0.2cm]
APOGEE ID & RA [$^\circ$] & Dec [$^\circ$] & $T_{\rm eff}$ [K] & $\log g$ & $v_{\rm micro}$ [km/s] & [C/H] & [N/H] & [O/H] & [Mg/H] & [Al/H] \\[0.1cm]
\tableline
\\[-0.2cm]
2M00000233+1452324 & 0.00975 & 14.87567 & 4809.8 & 4.42 & 1.62 & 0.21 & 0.59 & 0.33 & 0.37 & 0.54 \\
2M00000317+5821383 & 0.01323 & 58.36065 & 3744.9 & 0.95 & 1.68 & -0.44 & -0.05 & -0.29 & -0.10 & -0.12 \\
2M00000662+7528598 & 0.02762 & 75.48329 & 7322.1 & 4.11 & 2.05 & 0.05 & 0.05 & 0.04 & -0.23 & -0.27 \\
2M00011143+6240187 & 0.29765 & 62.67188 & 3839.4 & 1.08 & 1.77 & -0.48 & 0.26 & -0.25 & -0.08 & -0.03 \\
$\cdots$ & $\cdots$ & $\cdots$ & $\cdots$ & $\cdots$ & $\cdots$ & $\cdots$ & $\cdots$ & $\cdots$ & $\cdots$ & $\cdots$ \\
\tableline\\[0.2cm]

\tableline \tableline
\\[-0.2cm]
APOGEE ID & [Si/H] & [S/H] & [K/H] & [Ca/H] & [Ti/H] & [Cr/H] & [Mn/H] & [Fe/H] & [Ni/H] & [Cu/H] \\[0.1cm]
\tableline
\\[-0.2cm]
2M00000233+1452324 & 0.36 & 0.14 & 0.38 & 0.39 & 0.37 & 0.46 & 0.25 & 0.35 & 0.44 & 0.28 \\
2M00000317+5821383 & -0.12 & -0.21 & -0.37 & -0.34 & -0.20 & -0.23 & -0.39 & -0.28 & -0.20 & -0.38 \\
2M00000662+7528598 & 0.13 & 0.02 & -0.31 & -0.05 & -0.31 & -0.01 & -0.07 & -0.13 & -0.28 & 0.91 \\
2M00011143+6240187 & 0.03 & 0.03 & -0.21 & -0.23 & -0.08 & -0.19 & -0.28 & -0.15 & -0.17 & -0.27 \\
$\cdots$ & $\cdots$ & $\cdots$ & $\cdots$ & $\cdots$ & $\cdots$ & $\cdots$ & $\cdots$ & $\cdots$ & $\cdots$ & $\cdots$ \\
\tableline\\[0.2cm]
\end{tabular}

\begin{tabular}{lcccl}
\tableline \tableline
\\[-0.2cm]
APOGEE ID & $C_{12}/C_{13}$ & $v_{\rm macro}$ [km/s] & $\chi^2_R$ & Quality Flag \\[0.1cm]
\tableline
\\[-0.2cm]
2M00000233+1452324 & 51.5 & 1.02 & 9.9 & good \\
2M00000317+5821383 & 12.5 & 0.96 & 323.9 & chi2\_R > 50 \\
2M00000662+7528598 & 79.6 & 29.99 & 24.7 & vmac > 20kms \\
2M00011143+6240187 & 39.0 & 23.57 & 280.7 & vmac > 20kms ; chi2\_R>50 \\
$\cdots$ & $\cdots$ & $\cdots$ & $\cdots$ & $\cdots$ \\
\tableline\\[0.2cm]
\end{tabular}

\end{center}
\end{table*}

%
%
%
%
%
%

\section{Discussion}
\label{sec:discussion}

\tP\, provides a straightforward way to perform full spectral fitting with a minimal number of spectral models required; in our case, we only generated 2000 synthetic {\it ab initio} spectra for 25 stellar labels. \tP\, does not require a boutique spectroscopic mask \citep[e.g., APOGEE/ASPCAP,][]{gar16}, but only a simple spectroscopic mask, constructed algorithmically from the comparison of the synthetic and observed spectra of two standard stars. This appears to be sufficient to attain stellar labels that are more precise and broadly consistent with stellar evolution models. But it is important to emphasize that the main goal of this paper is to lay out this new fitting methodology, using APOGEE merely as a sample application. There are several limitations in the current APOGEE-{\it Payne} catalog.

Despite the improvement going beyond the quadratic models and a small median interpolation errors of $0.1\%$, the interpolation error can be larger than $1\%$ in some extreme cases (see Section~\ref{sec:testing-the-Payne}), and can still prohibit obtaining absolute abundances to the level better than $0.05-0.1\,$dex, especially for the cooler stars. Elements with only very weak and blended features may be more susceptible to the interpolation error, and the absolute abundances for individual stars could be biased upto $0.2\,$dex. Another limitation of this catalog is that we do not fit for stellar rotation, $v \sin i$, but rather adopt $v_{\rm macro}$ as an approximation. We found that for some hot stars with $T_{\rm eff} \gtrsim 6500\,$K, their $v_{\rm macro}$ values reach the boundary ($v_{\rm macro} = 30\,$km/s) of our training set and exhibit seemingly spurious abundance patterns. We create a flag in the catalog for these fast rotators and defer a proper accounting of $v \sin i$ to future studies; ultimately, this can just be another (two) labels to fit. Furthermore, as discussed in Section~\ref{sec:teff_and_logg}, there is a 100$\,$K inconsistency between our spectroscopic $T_{\rm eff}$ and external photometric $T_{\rm eff}$, which appears to favor more metal poor estimates at the high metallicity end and more metal rich estimates at the low metallicity end, with a discrepancy up to 0.1 dex. The reason for this discrepancy in unknown, but it seems to agree with the APOGEE uncalibrated $T_{\rm eff}$. It thus calls for a more careful analysis of the spectral models adopted in this study and H-band spectral models in general. We also truncate our training set at [Fe/H]$\,=-1.5$ and do not analyze metal poor globular clusters or halo stars. Further, we do not fit for unresolved stellar binaries that might affect the abundances for dwarf stars. Such an analysis was done separately in \citet{elb18b} using \tP. This illustrates that \tP\, is not a panacea for stellar spectroscopy --- is only a new methodological framework, and we mention a few areas for future improvement below.

Our analysis of the APOGEE also made simplifying assumptions about the experimental set-up. First, we convolve all synthetic spectra with a fixed averaged LSF template from APOGEE, assuming that the averaged LSF is an accurate representation for all APOGEE spectra. This is not the case because the instrumental dispersion can vary from fiber-to-fiber and observation-to-observation. In this application of \tP , we do not fit for the LSF, but use $v_{\rm macro}$ as a free parameter instead to compensate part of the LSF variation. Second, we normalize synthetic spectra the same way as we normalize observed spectra; but even with a self-consistent normalization, the normalization scheme can be still problematic at low temperatures. In particular, \citet{nes15a} derived a set of ``continuum pixels" for APOGEE giant spectra with $T_{\rm eff} \simeq 4$,$000-5$,$000\,$K. At lower temperatures, this set of pixels that we adopt might no longer be valid reference points, and the systematics between models and observations can still skew the continuum. In the long run, fitting the LSF and continuum along with the stellar labels might mitigate some of the remaining systematics seen in this study.

The success of \tP\, relies on further key ingredients. A robust spectroscopic mask, which we derived here from only the Sun and Arcturus. It will be crucial to have a set of standard stars that all large-scale spectroscopic surveys will observe: due to the subtle combined effect from the instrumental profile, telluric skylines, and normalization, we found that a robust mask must be made based on observations from the same instrument. It is, for example, not sufficient to make a spectroscopic mask using the highest resolution FTS spectra and convolve the mask to the observable space. Second, \tP\, must rely on {\it ab initio} spectral models that span a broad range of the $T_{\rm eff} - \log g - {\rm [Fe/H]}$ space. Again, a limitation of \tP 's current application derives from the fact that the line list underlying its {\it ab initio} models is only calibrated to two stars, the Sun and Arcturus. Models for cooler stars ($T_{\rm eff}<4,000$~K) and more metal poor stars therefore remain problematic. One future step will be extending the calibration of the line list beyond the Sun and Arcturus as well as constructing a spectroscopic mask beyond using these two stars. It is also essential to explore other more sophisticated options, such as 1D or 3D NLTE models. With \tP, only a few hundreds or thousands of spectral models are needed to fit for all stellar labels. Therefore, self-consistent NLTE elemental abundances should be possible with \tP\, (see Kovalev et al., {\it in prep.}).

%
%
%
%
%
%

\section{Summary}
\label{sec:conclusion}

We present \tP\, as a new framework for fitting stellar spectra with physical models which allows for the simultaneous determination of numerous stellar labels. The key ideas and components of \tP\, are summarized below:

\begin{enumerate}
\item \tP\, builds generative models that predict spectral fluxes at each pixel for any set of stellar labels, based on a modest discrete set of physical, {\it ab initio} model spectra. Only $\mathcal{O}(1000)$ suitable {\it ab initio} spectra are needed to create a model for fitting $N_{\rm dim} = 25$ labels. This opens up new avenues for determining stellar labels drawing on computationally expensive synthetic models, such as 3D NLTE models.

\item The generative models for each pixel are built with neural-networks-like function, a flexible extension of the quadratic models.

\item At any combination of stellar parameters and abundances, the underlying spectral models for the training step consistently calculate the atmosphere structure and the radiative transfer, to yield the {\it ab initio} spectra.

\item An auto-refinement technique iteratively determines the points in stellar label space where {\it ab initio} spectra are needed to construct a sufficiently high-quality generative model. 

\item For fitting of actual survey spectra, we apply a straightforward and algorithmic way to construct a pixel mask that excises wavelengths where observed spectra systematically and prominently differ from the model predictions.

\item The code for \tP\, is minimal and transparent, and the fitting is very efficient. Fitting 25 labels simultaneously to an APOGEE spectrum, and accounting for radial velocity at the same time, takes no more than a CPU second per spectrum. One significant advantage of this speed is the ability to explore many detailed options and choices in the fitting.
\end{enumerate}

\tP\, is a general method that can be applied to any set of synthetic models and survey data. As a sample application, we applied \tP\, to the full APOGEE DR14 catalog. By directly fitting all stellar labels simultaneously, \tP\, delivers the state-of-the-art results that appear both accurate and precise, without any external calibration. We demonstrate this by comparing the APOGEE labels derived from \tP\, to APOGEE DR14 results and subjecting them to a variety of physical plausibility tests, illustrating the following points:

\begin{enumerate}
\item Fitting all data only with a single generative model, we obtain stellar parameters ($T_{\rm eff}$, $\log g$, [Fe/H]) for both giants and dwarfs that are broadly consistent with external constraints, although modest [Fe/H]-dependent systematic offsets are evident when comparing to photometric temperatures and asteroseismic $\log g$ values.

\item We obtain detailed elemental abundances for APOGEE dwarf stars, which had not been done within APOGEE before. Broadly, the abundance patterns we see in the dwarfs are consistent with those in the giant sample.

\item The derived elemental abundances for stars within open clusters show no systematic trends with $T_{\rm eff}$.

\item We resolve the problem of the ``peculiar'' $[{\rm Ti/Fe}]-[{\rm Fe/H}]$ abundance distribution that has been persistent in APOGEE data releases and obtain a physically plausible abundance pattern.

\item Our derived [C/N] ratios agree well with predictions for surface abundance changes due to dredge-up in stellar evolution models. The agreement between stellar evolution models and our [C/N] ratios vividly demonstrates that it is possible to infer stellar ages directly from {\it ab initio} spectroscopic labels.

\item We achieve an elemental abundance precision of $\sim 0.03\,$dex for all elements. On this basis, we confirm that open clusters are chemically homogeneous to a level of at least $0.03\,$dex.
\end{enumerate}

\tP\, offers a very powerful framework to perform accurate, efficient comparison of large data sets to sophisticated spectral models. Work is currently underway to improve the training of \tP\, among the cool giants, to employ NLTE spectral models, and to apply \tP\, to a diverse variety of existing spectroscopic and photometric data sets.

%
%
%
%
%
%

\acknowledgments

We thank Robert Kurucz for developing and maintaining programs and databases and Fiorella Castelli for allowing us to use her linux versions of the programs without which this work would not be possible. We also want to thank Kareem El-Badry, Ben Johnson, Jieun Choi, Aaron Dotter, Martin Asplund, Ken Freeman, Mark Pinsonneault, David Weinberg, Jennifer Johnson, Jon Holtzman, Henrik Jonsson, David W. Hogg, Jo Bovy, Melissa Ness and Anna Ho for illuminating discussions. YST is grateful to be supported by the NASA Hubble Fellowship grant HST-HF2-51425.001 awarded by the Space Telescope Science Institute, and was supported by the Carnegie-Princeton Fellowship, the Martin A. and Helen Chooljian Membership from the Institute for Advanced Study in Princeton, the Australian Research Council Discovery Program DP160103747 and the NASA Headquarters under the NASA Earth and Space Science Fellowship Program - Grant NNX15AR83H for this project. C.C. acknowledges support from NASA grant NNX13AI46G, NSF grant AST-1313280, and the Packard Foundation. H.W.R.'s research contribution is supported by the European Research Council under the European Union's Seventh Framework Programme (FP 7) ERC Grant Agreement n.$\,$[321035].

%
%
%
%
%
%

\end{CJK*}

\vspace{1cm}

\bibliography{biblio.bib}

\end{document}